\documentclass[journal=jacsat,manuscript=article,preprint
]{achemso}

\usepackage[version=3]{mhchem} 
\usepackage{color}
\usepackage{soul}

\author{Emanuele Coccia}
\affiliation[University of Padova]
{Department of Chemical Sciences, University of Padova, via Marzolo 1, Padova, Italy}
\alsoaffiliation{S3 Center, CNR Institute of Nanoscience, via Campi 213/A, Modena, Italy} 
\email{emanuele.coccia@unipd.it}
\author{Filippo Troiani}
\affiliation[CNR]{S3 Center, CNR Institute of Nanoscience, via Campi 213/A, Modena, Italy} 
\author{Stefano Corni}
\affiliation[University of Padova]
{Department of Chemical Sciences, University of Padova, via Marzolo 1, Padova, Italy}
\alsoaffiliation{S3 Center, CNR Institute of Nanoscience, via Campi 213/A, Modena, Italy}
\email{stefano.corni@unipd.it}

\SectionNumbersOn 

\title[Quantum coherence in molecules]
  {Probing quantum coherence in ultrafast molecular processes: an ab initio approach to open quantum systems}

\keywords{American Chemical Society, \LaTeX}

\begin{document}







\begin{abstract}

Revealing possible long-living coherence in ultrafast processes allows detecting genuine quantum mechanical effects in molecules. To investigate such effects from a quantum chemistry perspective, we have developed a method for simulating the time evolution of molecular systems, based on {\it ab initio} calculations that includes relaxation and environment-induced dephasing of the molecular wave function, whose rates are external parameters. 
The proposed approach combines a quantum chemistry description of the molecular target with a real-time propagation scheme within the time-dependent stochastic Schr\"odinger equation. Moreover, it allows a quantitative characterization of the state and dynamics coherence, through the $l_1$-norm of coherence and the linear entropy, respectively. To test the approach, we have simulated femtosecond pulse-shaping ultrafast spectroscopy of terrylenediimide, a well studied fluorophore in single-molecule spectroscopy. 
Our approach is able to reproduce the experimental findings [R. Hildner {\it et al.},Nature Phys., {\bf 7}, 172 (2011)], confirming the usefulness of the approach and the correctness of the implementation.
\end{abstract}

\section{Introduction}

\begin{figure}[tbp]
\includegraphics[width=0.85\textwidth]{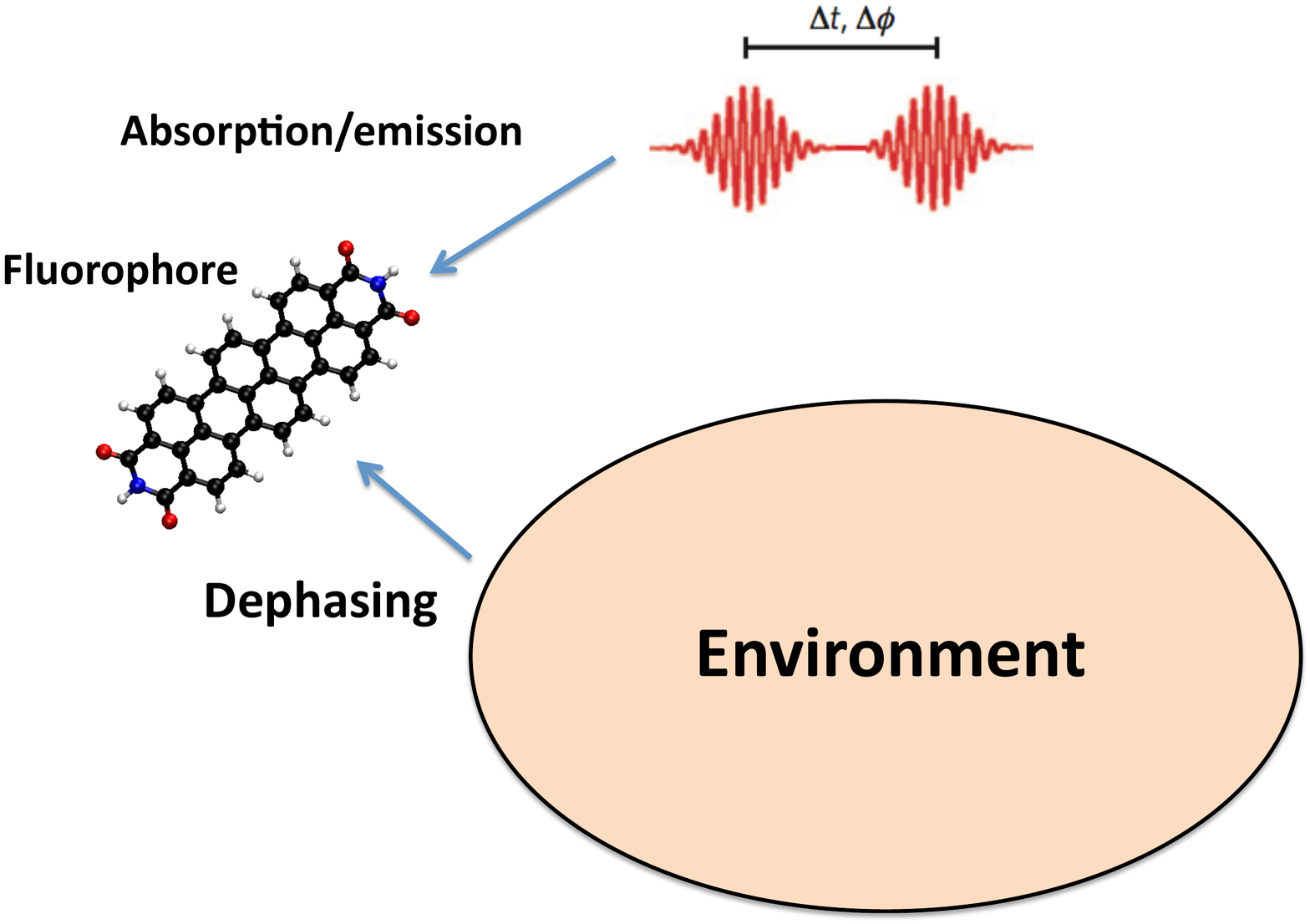}
\caption{Schematic representation of the system: the fluorophore interacts with a sequence of two pulses, with varying time delays $\Delta t$ and phase shifts $\Delta \phi$, in presence of an external environment leading to dephasing of the molecular wave function. Ball-and-stick representation of TDI is shown. \label{fig1} }
\end{figure}

Ultrafast spectroscopy is a powerful tool to investigate, control and manipulate quantum coherence in molecules and complex systems \cite{eng07,lee07,ish09,sar10,ish12,cer03,pol10,roz13,fal14,cer17,sto17,muk00,abr09,col10,kim09,sch17,tdi,bri10,hil13,bri14}. Detection of electronic and vibrational coherence in biological systems, as light-harvesting complexes involved in photosynthesis, and interpretation of the experimental evidences, is a matter of stimulating and open debate \cite{gro06,sch10,col13,hil13,che15}. Recently, specific ultrafast spectroscopy techniques could probe a single molecule \cite{tdi,bri10,hil13,bri14}. To understand the outcomes of such experiments, theoretical and computational approaches are required, able to include all the important features of the simulated system. One such feature is certainly the ubiquitous coupling with the environment, that qualifies the probed system as {\it open}. In principle, every system must be considered as open, namely interacting with a surrounding, larger environment \cite{bk:open,bk:open1,bk:open2,bk:open3,bk:open4}. 
Abstraction process leading to closed and isolated systems can indeed represent a crude approximation of the intrinsic features of the microscopic world.
For this reason, coupling between quantum systems and an external environment is essential to give a complete description of physical phenomena \cite{sch05,dal14}, also in the ionization regime \cite{tre11}. \\
Time-dependent Schr\"odinger equation (or, equivalently, von Neumann equation for the time evolution of the density matrix) describes a coherent dynamics, i.e. a dynamics with a well defined phase relation between the eigenstates of the system. 
In principle, one could extend the boundary of the system and include the environment in a larger system, but this is typically impractical computationally
due to the enormous number of degrees of freedom involved in describing the environment \cite{sse1}. Moreover, one usually is not interested in the microscopic description of the environment, which in most cases can be regarded as an external bath. \\
A feasible way to treat open systems is to reduce the number of degrees of freedom, by tracing out those of the bath, and defining the so-called reduced density matrix $\hat{\rho}_S$ \cite{bk:open,bk:open1,bk:open2,bk:open3,bk:open4,lind76,red,ego08,che16}.
Assuming that bath relaxation timescales are much faster than those of the system \cite{sse1}, i.e. the bath is seen unchanged from the system standpoint, the time evolution of the molecular wave function is not affected by memory effects. This corresponds to the Markovian limit, an approximation used in this work\cite{sse1,hak72,hak73}. \\
Supposing a weak coupling between system and bath, and that bath degrees of freedom are in thermal equilibrium at any time, one obtains the  Lindblad master equation for $\hat{\rho}_S$ (in the Markovian limit)\cite{lind76,gor76}. An alternative and less explored approach to the same problem is given by the Markovian stochastic Schr\"{o}dinger equation (SSE). Within SSE, one directly follows the time evolution of the system wave function, $\vert \Psi_S(t) \rangle$, in presence of dissipation and fluctuation effects induced by the environment. SSE approach overcomes lack of microscopic knowledge about the environment by including stochastic terms in any single realization of the wave function evolution. In short, SSE time propagation of the system wave function has been seen to be fully equivalent to solve Lindblad equation for $\hat{\rho}_S$, in the limit of infinite number of quantum trajectories \cite{sse1}. The main computational advantage of using SSE lies in the fact that the system wave function only depends linearly on the number of states of the system $N_{\text{states}}$, while $\hat{\rho}_S$ shows a quadratic dependence. On the other hand, although the single SSE realization is characterized by a linear dependence on $N_{\text{states}}$, a large number of trajectories has to be produced: this issue can be tackled computationally by exploiting the inherent parallel nature of the procedure. \\
Several theoretical and computational protocols for the numerical propagation of SSE have been defined over the years \cite{qjump1,qjump2,qjump3,revqjump,hof13,hig01,mak99,gis93}. Details on our choice will be given below. \\
It is worth to mention that well-established approaches are present in literature, that include non-Markovian effects, and make less restrictive assumptions about the quantum nature of the dynamics \cite{tan89,mak94,mak95,yan05,sch06,rod09}. The hierarchical equation of motion method\cite{tan89,yan05,sch06,rod09}, for instance, is based on a hierarchy of auxiliary density matrices to account for non-Markovian dynamics.
In the context of pigment-protein complexes approaches based on theory of non-Markovian open quantum systems has been successfully combined with QM/MM techniques for the study of quantum effects in photosynthesis \cite{dam02,olb11,shi12,val12,riv13,wan15}. \\
In this work we present a computational approach based on an ab initio description of fluorophores, that includes the effect of dephasing and relaxation via the Markovian SSE. In particular, we show how this method can be applied to simulate and interpret ultrafast spectroscopy measurements.  Coupling SSE to an {\it ab initio} representation of the electronic Hamiltonian of the molecular target is the key ingredient of the approach proposed here, allowing us to define dephasing and relaxation effects in the Hilbert space defined by the specific quantum-chemistry method adopted. \\
 A step forward in the definition of a computational protocol based on SSE is also the application of global and rigorous quantifiers to analyse coherence in (bio)chemical and physical systems. A growing interest on the development of a systematic theory of quantum coherence as a physical resource has recently arisen \cite{str17}. By reconstructing the density matrix from the ensemble of SSE trajectories, we can investigate the time evolution of quantum coherence by means of well established quantifiers, as $l_{1}$-norm \cite{bau14} and linear entropy \cite{zur93}. A quantitative analysis of quantum coherence is therefore possible, starting from an {\it ab initio} description of the molecular target. Enabling such kind of analysis, common for model Hamiltonians but rather original in an {\it ab initio} framework \cite{roz17}, is an important result of the  present work. \\
We have included in the real-time model developed recently \cite{pip16} 
the dephasing due to the environment surrounding the molecule and the relaxation, i.e., the spontaneous decay from the excited states to the ground state. The latter may effectively simulate non-radiative decay due, e.g., to internal conversion or radiative decay. Time-dependent $\vert \Psi_S(t) \rangle$ is expanded into the set of time-independent eigenstates of the fluorophore, obtained in this work at Configuration Interaction with singly excited configurations (CIS) \cite{tre08,her15,kli16}, with a perturbative correction for energies involving doubly excited configurations (CIS(D)).
Only few alternative examples of coupling an {\it ab initio} description with a treatment of dephasing and relaxation are present in literature. Most of them use {\it ab initio} results as input for master-equation approaches \cite{zen12,riv14}. Stochastic approaches have been used in the context of TDDFT, at various level of complexity \cite{dag08,app11,tem11,hof13}. The proposed approach is alternative to both, avoiding the possible artifacts of TDDFT beyond the linear regime \cite{rag11,rag12,rag12b,hab14} and defining a seamless integration of {\it ab initio} and open-quantum systems approaches. \\
As a realistic test case, we have considered pulse-shaping spectroscopy on the terrylenediimide (TDI) fluorophore \cite{tdi} (Figure \ref{fig1}). The authors of the original work reported how to control  and manipulate electronic coherence in single molecules, by studying the interplay between variations of coherence and emission of TDI. The choice of TDI to validate our approach gives us the opportunity to compare our results with well established experimental findings. \\
The paper is organized as follows. In Section \ref{theory}, first SSE is briefly reviewed, then we discuss how relaxation and pure dephasing channels \cite{tem11a} are introduced in the chosen quantum chemistry framework (CIS) using the version of quantum jump algorithm proposed in Refs. \cite{qjump1,qjump2}, and finally we provide the definition of the quantum coherence quantifiers used in this work. The numerical results are shown and discussed in Section 3, while conclusions and perspectives are collected in Section \ref{con}.

\section{Theory}
\label{theory}

\subsection{Stochastic Schr\"odinger equation} 

As reported in the Introduction, the fluorophore interacts with an environment causing dephasing in the wave function. In the theory of open quantum systems \cite{bk:open}, the total Hamiltonian $\hat{H}$ is defined as the sum of the Hamiltonian of the system (S), of the bath (or environment, B) and their mutual interaction:
\begin{eqnarray}
\hat{H}(t) & = & \hat{H}_{S}(t)  \otimes \hat{I}_{B} + \hat{I}_{S} \otimes \hat{H}_{B} + \alpha \hat{H}_{SB} \\ 
 \hat{H}_{SB} &=& \sum_{q}^{M} \hat{S}_{q} \otimes \hat{B}_{q},
 \label{eq:start}
\end{eqnarray}
with $\hat{I}_S$ and $\hat{I}_B$ being the identity in the system and bath Hilbert space, respectively. In the present study, the system is given by the fluorophore. The  strength factor $\alpha$ modulates the interaction between the system and the bath. The interaction term $\hat{H}_{SB}$ has a bilinear form, characterized by the two sets of operators $\hat{S}_q$ and $\hat{B}_q$, operating on the Hilbert space of the system and of the bath, respectively. Operator $\hat{S}_{q}$  describes the effect of the bath on the system, while the operator $\hat{B}_q$ describes how the bath is affected by the presence of the system. The sum in $\hat{H}_{SB}$ runs over the number $M$ of interaction channels $q$ between the system and the bath.
Since our final goal is to simulate optical processes, in $\hat{H}_{S}(t)$ the time-dependent interaction term with the external (classical) electromagnetic field is also included 
\begin{equation}
\hat{H}_{S}(t) = \hat{H}_{S}^0 - \hat{\vec{\mu}} \cdot \vec{E}(t),
\end{equation}
where $\hat{H}_{S}^0$ is the Hamiltonian of the isolated system and $\hat{\vec{\mu}}$ is the system dipole interacting with the external electric field $\vec{E}(t)$.
 \\
In this work, only the Markovian limit will be explicitly taken into account, corresponding to a delta approximation of the time autocorrelation function of the bath, on the timescale of the system response \cite{sse1}. In other words, the loss of information from the system is irreversible \cite{pii08}.
As pointed out in the Introduction, since the resolution of the full problem defined in Eq. 1 is impractical for realistic cases, the total number of degrees of freedom is drastically reduced when those of the bath are traced out from the total (i.e. $S+B$) density matrix $\rho$: the effect of the bath is treated in an effective way by introducing the reduced density matrix $\hat{\rho}_{S}$ of the system (time dependence is made explicit):
\begin{equation}
\hat{\rho}_{S}(t) = \text{Tr}_{B} \hat{\rho}(t).
\end{equation}
Under the assumption of Markovian behaviour and of weak coupling between system and (up to the second order in the interaction term $\alpha$), the time evolution of the reduced density matrix $\rho_S$ follows the Lindblad master equation \cite{lind76}:
\begin{eqnarray}
\frac{d}{dt} \hat{\rho}_{S}(t) &=& -i [\hat{H}_{S}(t),\hat{\rho}_{S}(t)] +  \hat{\mathcal{L}} \hat\rho_S(t) \\ 
\hat{\mathcal{L}} \hat\rho_S(t) &=& -\frac{1}{2} \sum_{q}^{M}\{\hat{S}_{q}^{\dagger}\hat{S}_{q}, \nonumber \hat{\rho}_{S}(t)\} + \sum_{q}^{M} \hat{S}_{q}\hat{\rho}_{S}(t)\hat{S}_{q}^{\dagger}. \\ \nonumber
\label{eq:lind}
\end{eqnarray}
The first term in the rhs of Equation 5 corresponds to the von Neumann time-evolution of the density matrix of a closed system. The interaction with the bath is described by the Lindblad superoperator $\hat{\mathcal{L}}$.
The analytical solution of Equation 5 is known only for few model systems, otherwise a numerical real-time propagation is mandatory. The main drawback of solving the Lindblad equation lies in the dimension of reduced density matrix, i.e. $N_\text{states}^2$, which could make numerical simulations computationally demanding. \\
An alternative but  equivalent approach is given by the stochastic Schr\"odinger equation (SSE) \cite{sse1}, written in the Markovian limit as:
\begin{equation}
i \frac{d}{dt} |\Psi_{S}(t) \rangle = \hat{H}_{S}(t) |\Psi_{S}(t) \rangle + \alpha \sum_{q}^M l_{q}(t) \hat{S}_{q} |\Psi_{S}(t) \rangle - \alpha^{2} \frac{i}{2} \sum_{q}^M \hat{S}_{q}^{\dagger}\hat{S}_{q} |\Psi_{S}(t) \rangle.
\label{eq:sse}
\end{equation}
Starting from the fluctuation-dissipation theorem \cite{bk:kam}, one can interpret  the nonHermitian term $- \alpha^{2} \frac{i}{2} \sum_{q}^M \hat{S}_{q}^{\dagger}\hat{S}_{q}$ as the dissipation due to the environment, whereas $\sum_{q}^M l_{q}(t) \hat{S}_{q}$ is the fluctuation term, modeled by a Wiener process $l_{q}(t)$, i.e. a white noise associated to the Markov approximation. \\
 The main advantage in using SSE is that one directly treats the system wave function, that only depends linearly on $N_\text{states}$, thus saving computational time with respect to the propagation of $\hat{\rho}_S(t)$.
Diagonal and off-diagonal elements of the reduced density matrix $\hat{\rho}_S(t)$, respectively populations and coherences of the states of the system at time $t$, are obtained  by averaging on the number of independent realizations $N_{\text{traj}}$ of propagating SSE. Given the definition of the reduced density matrix 
\begin{equation}
\hat{\rho}_S(t) \equiv \frac{1}{N_{\text{traj}}} \sum_j^{N_{\text{traj}}} \vert \Psi_{S,j}(t) \rangle \langle \Psi_{S,j}(t) \vert,
\end{equation}
where $|\Psi_{S,j}(t) \rangle$ is the system wave function corresponding to $j$-th realization, and expanding $\vert \Psi_{S,j}(t) \rangle$ into stationary eigenstates $| \Phi_m \rangle$ of the system
\begin{equation}
|\Psi_{S,j}(t) \rangle = \sum_m^{N_\text{states}} C_{m,j}(t) | \Phi_m \rangle,
\label{eq:cis}
\end{equation}
one defines population and coherences as the following:
\begin{eqnarray}
\text{population of state} \quad q & \equiv & (\hat{\rho}_S(t))_{qq} = \frac{1}{N_{\text{traj}}} \sum_j^{N_{\text{traj}}} \vert C_{q,j}(t) \vert^2 \\
\text{coherence of states} \quad q \quad \text{and} \quad k & \equiv& (\hat{\rho}_S(t))_{qk} = \frac{1}{N_{\text{traj}}} \sum_j^{N_{\text{traj}}} C_{q,j}^*(t) C_{k,j}(t).
\end{eqnarray}
In the limit of large number of quantum trajectories, SSE reproduces the same $\hat{\rho}_S(t)$ coming out from the Lindblad master equation \cite{sse1}. \\
 For sake of clarity, we will use in the following the definition: $\hat{H}_{SSE}(t) \equiv \hat{H}_{S}(t) + \alpha \sum_{q}^M l_{q}(t) \hat{S}_{q} - \alpha^{2} \frac{i}{2} \sum_{q}^M \hat{S}_{q}^{\dagger}\hat{S}_{q} $, with $\alpha=1$.

\subsection{CIS expansion of the wave function}

 SSE, which has been used in the past for model (typically two-state) systems\cite{rit15,zha16}, is here coupled to a quantum-chemistry description of the molecular target. In this work,  in the expansion of Eq. \ref{eq:cis} 
$C_m(t)$ are time-dependent expansion coefficients, and $| \Phi_m \rangle$ represents the $m$-th time-independent CIS eigenstate of the isolated system, with eigenvalue $E_m$. 
The CIS eigenstates are symmetry-adapted linear combinations of singly-excited Slater determinants \cite{cis}:
\begin{equation}
| \Phi_m \rangle = d_0 | \psi_0 \rangle + \sum_i^{\text{occ}} \sum_a^{\text{vir}} d_{i,m}^a | \psi_i^a \rangle, 
\end{equation}
where $| \psi_0 \rangle$ is the reference HF state, $| \psi_i^a \rangle = \hat{a}^{\dagger}_a \hat{a}_i | \psi_0 \rangle$ is the
configuration obtained by the single excitation from the occupied HF
orbital $i$ to the virtual HF occupied $a$, and the coefficients
$d_0$ and $d_{i,m}^a$ are obtained by diagonalizing the Hamiltonian of the isolated molecule in this space. Each Slater determinant is defined as the antisymmetrized product of single-electron molecular orbitals $\phi_l(\textbf{r})$ 
\begin{equation}
\phi_l(\textbf{r}) = \sum_\mu^{N_{\text{basis}}} \lambda_{l\mu} \chi_\mu(\textbf{r}),
\end{equation}
expanded on $N_{\text{basis}}$ Gaussian basis functions, where $\textbf{r}$ is the collective electronic coordinate. \\
However, our model is general and not limited to a CIS expansion of the wave function. 
Our computational protocol is therefore articulated in two main steps: first, a quantum-chemistry calculation; second, a real-time SSE propagation with these ab initio quantities. \\ 
The matrix form of the SSE is formally given by
\begin{equation}
\frac {\partial \textbf{C}(t)} {\partial t} = \textbf{H}_{SSE}(t) \textbf{C}(t), 
\end{equation}
where $\textbf{C}(t)$ is the vector of the time-dependent expansion coefficients and $\textbf{H}_{SSE}(t)$ is the matrix representation at time $t$ of $\hat{H}_{SSE}(t)$ in the basis of the CIS eigenstates ($\textbf{H}_{SSE}(t))_{qk} = \langle\Phi_q | \hat{H}_{SSE}(t) | \Phi_k \rangle$. 

\subsection{Including relaxation decay}


Relaxation refers to the decay from an electronic excited state of the fluorophore to its ground state $\vert \Phi_0 \rangle$ \cite{tre08}. It can be due to photon emission (radiative decay) or to nonradiative decay (e.g. through internal conversion). The former can be seen as an effect of the electromagnetic field seen as an environment, the latter is more molecular based, although the ladder of vibrational levels may also be seen as an environment for the electronic level. \\
Nonadiabatic vibronic coupling provides a nonradiative decay channel, which plays an essential role in the molecular relaxation process\cite{ton01,san07}. In the SSE framework, such process is accounted by the operator

\begin{equation}
 \hat{S}_q^{\text{rel}} = \sqrt{\Gamma_q} | \Phi_0 \rangle \langle \Phi_q |,
 \label{eq:rel}
\end{equation} 

 which induces an exponential decay of the population $|C_q(t)|^2$ and quantum jumps corresponding to the collapse of the system wavefunction $\vert \Psi_S(t) \rangle$ into the ground state $|\Phi_0\rangle$. The nonradiative relaxation rate $\Gamma_q$ can be regarded as a phenomenological parameter or obtained by ab initio nonadiabatic simulations \cite{cur18}. The same operator can be used to account for the radiative relaxation. Here, given the matrix element of the optical transition $q \rightarrow 0$, the decay rate is derived through the Fermi's Golden Rule. In the presence of both decay channels, the overall decay rate is given by the sum of the radiative and nonradiative ones.



\subsection{Including pure dephasing}

Dephasing acts on the decay of the off-diagonal elements of the reduced density matrix, i.e. the coherences of the system. The choice of the form of the dephasing operator is not univocal. Dephasing operators are usually given for two-state (2s) systems as proportional to the $\sigma_z$ Pauling matrix
\begin{equation}
\hat{S}^{\text{dep}}_{\text{2s}} = \sqrt{\gamma_{\text{2s}}/2} \left(| \Phi_1 \rangle \langle \Phi_1 | - | \Phi_0 \rangle \langle \Phi_0 | \right),
\end{equation}
with $\gamma_{2s}$ the associated dephasing rate.\\
Here, we extend this form to a generic multi-state system. In fact, we define an operator for the pure dephasing that changes the sign of the element $|\Phi_q \rangle \langle \Phi_q | $; more specifically, it is
\begin{equation}
\hat{S}_{q}^{\text{dep}} = \sqrt{\gamma_{q}/2} \sum_p^{N_\text{states}}  M(p,q) | \Phi_p \rangle \langle \Phi_p |,
\label{eq:dep}
\end{equation}
where $M(p,q)$ is equal to -1 if $p = q$ or equal to 1 otherwise. 
The operator in Eq. \ref{eq:dep} guarantees that the population of the various states remains unchanged during the propagation (and not only on average), i. e. $\vert C_q(t_j) \vert^2 = \vert C_q(t_0) \vert^2$ for each $j>0$, with $t_0$ being the initial time. Moreover, the specific definition of the dephasing operator in Eq. \ref{eq:dep} keeps the population unchanged also in presence of  a quantum jump. Off-diagonal elements of the reduced density matrix $(\hat{\rho}(t))_{qk}$ exponentially decay with a rate equal to $\gamma_q + \gamma_k$; this results directly comes out from the analysis of $\hat{S}_{q}^{\text{dep}}$ and $\hat{S}_{k}^{\text{dep}}$ in the Lindblad superoperator. $\gamma_q$ and $\gamma_k$ are introduced as phenomenological parameters, since treating dephasing at ab initio level is challenging, due to the interaction between the system and the many degrees of freedom of the environment. \\
Given the definitions of the operators in Eqs. \ref{eq:rel} and \ref{eq:dep}, one can verify that the number of interaction channels $M$ coincides with $N_{\text{states}}$.

\subsection{Quantum jump algorithm}

SSE propagation can be performed by means of a quantum jump algorithm \cite{qjump1,qjump2,qjump3,revqjump,hof13}: a deterministic nonHermitian dynamics is coupled to a number of random jumps (simulating fluctuation induced by the bath), obtained with Monte Carlo techniques \cite{mak99}. Alternatively, one can use the quantum state diffusion model \cite{gis92,gis93}, in which the propagation is performed in terms of continuous stochastic differential equations, in linear or nonlinear form \cite{dag08}. 
Continuous propagators \cite{hig01}, as the Euler-Maruyama \cite{hig01} or the Leihmulker-Matthews \cite{lei13,lei13b} methods, can be used. \\
In the present investigation we have used the quantum jump algorithm, as proposed in Ref. \cite{qjump2}. The Hamiltonian $\hat{H}_{nH}(t)$ of the deterministic nonHermitian part of the real-time propagation is given by
\begin{equation}
\hat{H}_{nH}(t) = \hat{H}_S(t) - \frac{i}{2} \sum_{q}^M \hat{S}_{q}^{\dagger}\hat{S}_{q}.
\end{equation}
A second-order version of the Euler algorithm is used to propagate the coefficients ${\textbf{C}}(t)$ \cite{pip16,pip17} of the wave function expansion, leading to
\begin{equation}
\textbf{C}(t+\delta t) = \textbf{C}(t-\delta t) -2i\delta t \textbf{H}_{nH} \textbf{C}(t),
\end{equation}
where $\delta t$ is the finite time step used for the numerical propagation of the deterministic part of SSE. 
The time evolution based on $\hat{H}_{nH}(t)$ does not conserves the norm of the wave function. \\
At first order in $\delta t$, the norm $\eta$ of the time-dependent wave function $| \Psi_S(t) \rangle $ at the time $t_{j+1} = t_{j} + \delta t$ can be written as
\begin{equation}
\eta = \langle \Psi_S(t_{j+1}) | \Psi_S(t_{j+1}) \rangle = 1 - \Delta p
\end{equation}
with
\begin{eqnarray}
\Delta p &=& i \delta t  \sum_q^M \Delta p_q \\
\Delta p_q &=& \langle \Psi_S(t_j) | \hat{S}_q^\dagger \hat{S}_q | \Psi_S(t_j)  \rangle.
\end{eqnarray}
The quantity $\Delta p$ represents the probability that a generic quantum jump occurs, while $\Delta p_q$ defines the probability that the quantum jump involves the specific interaction channel $q$, of relaxation or dephasing type. Both probabilities are imposed using Monte Carlo techniques.
In detail, $\Delta p$, which therefore corresponds to the amount of lost norm in the time step $\delta t$ from time $t_j$ to $t_{j+1}$, is compared at each step with a random number $\epsilon$ uniformly distributed between 0 and 1:
\begin{itemize}
\item
if $\Delta p < \epsilon$ no quantum jump occurs, and the wave function is then normalized;
\item
if $\Delta p \ge \epsilon$, a quantum jump occurs, and the new function is defined as the following
\begin{equation}
| \Psi_S(t_{j+1}) \rangle = \frac{\hat{S}_q | \Psi_S(t_j) \rangle} {\sqrt{\Delta p_q / \Delta t}}
\end{equation}
with probability $ \frac{\Delta p_q} {\Delta p}$, determined using again the same Monte Carlo technique. 
\end{itemize}
Within the CIS expansion of $|\Psi_S(t) \rangle$ and using the relaxation and dephasing operators defined in Eqs \ref{eq:rel} and \ref{eq:dep}, one finds for the relaxation channel:
\begin{eqnarray}
\Delta p &=&  \delta t \sum_q |C_q(t)|^2 \Gamma_q \\
\label{eq:pop1}
\Delta p_q &=&  \delta t |C_q(t)|^2 \Gamma_q 
\label{eq:pop2}
\end{eqnarray}
and for the dephasing one:
\begin{eqnarray}
\Delta p &=&  \delta t \sum_q \gamma_q/2 \\
\Delta p_q &=&  \delta t \gamma_q/2.
\label{eq:coh2}
\end{eqnarray}
Clearly Eq. \ref{eq:pop2} corresponds to an exponential decay of populations, as anticipated before. Consequences of Eq. \ref{eq:coh2} are less obvious, but still an exponential decay of the coherences is obtained. The dephasing operator defined in Eq. \ref{eq:dep} maintains the population unchanged for each trajectory because $(\hat{S}_{q}^{\text{dep}})^{\dagger} \hat{S}_{q}^{\text{dep}}$ is equal to the identity. 

\subsection{Quantifying coherence}

As reported in the Introduction,  different quantifiers of quantum coherence have been introduced in the last years \cite{str17}, fulfilling a number of fundamental requirements. 
 In the following, we refer to the so-called $l_1$-norm of coherence, defined as \cite{bau14}
\begin{equation}
C_{l_1}(\hat{\rho}_S(t)) = \sum_{q \ne k} \vert (\hat{\rho}_{S}(t))_{qk} \vert,
\label{l1}
\end{equation}
which is time-dependent in our simulations due to the time evolution of the reduced density matrix. One can show that $C_{l_1}$ varies from zero for a fully incoherent state to $d$-1 (where $d$ is the dimension of the Hilbert space) for a maximally coherent state.
This quantity corresponds to the smallest distance, as quantified by the $l_1$-norm (i.e., the least absolute deviation), between the density matrix operator at a given time and that of any incoherent state, and describes the wavelike character of the state of the system.
The $l_1$-norm coherence has an intuitive interpretation, since it is given by the sum of the moduli of the off-diagonal elements of the reduced density matrix $\hat{\rho}_S$, which is built from the ensemble of SSE trajectories.  \\
The second quantifier used here is the linear entropy \cite{zur93}
\begin{equation}
S_L(\hat{\rho}_S) = 1 - \text{Tr}(\hat{\rho}_S^2),
\label{le}
\end{equation}
which refers to another interpretation of coherence, related to the system dynamics. Dynamics is indeed defined coherent if the time evolution at any time $t$ $>$ 0 can be expressed as a
unitary transformation of the initial state. If the system
is initialized in a pure state, typically the ground state,
it remains in a pure state in the presence of a coherent
evolution. The degree of coherence of the dynamics can
thus be quantified in terms of the purity of the density
operator throughout the time evolution. 
The trace of $\hat{\rho}^2$
varies from 1, for pure states, to 1/$d$, for a maximally
mixed state. Correspondingly, the linear
entropy varies from 0 to 1 - 1/$d$. 

 \section{Numerical tests: application to fs pulse-shaping spectroscopy on TDI}

In this section we consider fs pulse-shaping spectroscopy on TDI, simulated by means of the theoretical model described above. The goal is to validate the proposed approach by reproducing the experimentally observed emission properties of TDI \cite{tdi}. The latter is interrogated with a sequence of two pulses (Figure 1): the first pulse generates a coherent superposition of ground and first excited state, the second one (switched on with a given time delay $\Delta t$) probes the phase memory in TDI. In the following, we report both the values of the excited-state population, which is proportional to the observed fluorescence signal \cite{tdi}, and the time evolution of the off-diagonal matrix elements. The interest in the coherences is twofold: on one hand, they are responsible for the interference between the excitations produced by the two pulses; on the other hand, the coherences are related to the wavelike character of the system state, and thus its "quantumness" as quantified, e.g., by the $l_1$ norm of coherence and by the linear entropy. These quantities are computed as a function of the time delay $\Delta t$ and of the phase shift $\Delta \phi$ between the two pulses, for several dephasing times $T_2$. We have also explored the effect of the detuning $\delta$, corresponding to the difference between the central frequency of the pulsed field and the energy of the molecular transition between the ground and the first excited state of TDI, which turns out to be the most relevant one in the present case. \\
 TDI has been extensively studied from the experimental side \cite{tdi}.  Our aim is to test the approach reported above on the detection of quantum coherence of this fluorophore by reproducing the two-pulse spectroscopy results. 
 
\subsection{Computational details}
\label{det}

The geometry of the TDI molecule has been optimized at the DFT level, with the B3LYP functional and the 6-31G(d) basis set.
Time-independent CIS and CIS(D) calculations on the optimized TDI structure have been carried out using a locally modified version of Gamess \cite{gam1,gam2}. A 6-31G(d) basis set has been employed;
10 excited states have been kept in the expansion of the time-dependent wave function, corresponding to excitation energies up to 5 eV. { CIS and CIS(D) excitation energies are collected in Table S1 of Supporting Information.} A nonradiative decay time of 3.5 ns \cite{tdi} and a pure dephasing time of 30, 60 and 120 fs have been chosen. { The nonradiative decay time is taken from the experimental work \cite{tdi}, and we have selected the three values of dephasing time from the experimental distribution \cite{tdi}: 60 fs represents the maximum of the distribution, while 30 and 120 fs represent the extreme values detected in the experiment. All these values are input parameters in our model.} Detuning values of 80 and 160 cm$^{-1}$ have been employed in some of the simulations.  \\
For all the cases studied here, dynamics of 1 ps have been considered. A time step $\delta t$ of 1.21 as has been employed in all the simulations.  \\
In general, any pulse shape could be chosen, since pulses are encoded numerically. In this case, the two pulses are shaped with a Gaussian envelope function 
{
\begin{equation}
\vec{E}(t)= \vec{E}_{max} \exp \left   (  - \frac{(t-t_{0})^{2}}{2\sigma^{2}}   \right ) \sin(\omega t) + \\  
                   \vec{E}_{max}  \exp   \left   (  - \frac{(t-t_{0}-\Delta t)^{2}}   {2 \sigma^{2}}   \right )  \sin(\omega t + \Delta \phi) 
\end{equation}
where $\vec{E}_{max}$ is the maximum field amplitude, $t_0$ is the center of the first pulse, $\sigma$ is the width of the Gaussian  and $\omega$ the carrier frequency.
FWHM has been chosen to be equal to 49 fs, corresponding to an energy bandwidth of 48 cm$^{-1}$. The most part of the calculations has been carried out  with an  intensity I=${\frac{1}{2} \epsilon_0 c \vert \vec{E}_{max} \vert }^2$=5x10$^3$W/cm$^2$, while for a direct comparison with the experimental results we have also used I=6.6x10$^8$ W/cm$^2$. }
The wavelength is equal to 501 nm, coinciding with the CIS(D) transition from the ground $| \Phi_0 \rangle$ to the first excited state $| \Phi_1 \rangle$. 
Time delays of 0, 10, 30, 50, 100, 200, 400 and 600 fs have been used, together with  a phase shift $\Delta \phi$ of 0 and $\pi$. \\
We have used 512 quantum trajectories for each of the simulations reported in this paper: this number assured acceptable statistical errors. { In Figure S1 of the Supporting Information the convergence of the SSE results with respect to the number of trajectories is reported. }{ Quantum jumps associated to the relaxation do not occur along the 1 ps dynamics. SSE propagation with only dissipation due to the relaxation (while quantum jumps associated to the dephasing have been observed in our simulations) produces the right time evolution of the system wave function, which refers to the first excited state as an example. Since the absolute value of $C_1(t)$ is much smaller than 1, the expected exponential decay is indeed recovered without intervention of quantum jumps. Indeed, at first order in $\delta t$ and only considering dissipation from relaxation (no jumps) one obtains 
\begin{equation}
C_1(t+\delta t)=C_1(t) (1-\Gamma_1 \delta t (1-|C_1(t)|^2)).
\label{fo}
\end{equation}
}
The real-time propagation of the wave function, with the addition of relaxation and dephasing through SSE, has been performed using the homemade WaveT code \cite{pip16}.

\subsection{Results}
\label{res}
 
In the following, we show the simulated time evolution of the system state, and analyze in some detail both the excited-state populations and the coherence between ground and the first excited state. 
In order to verify the reliability of the proposed approach and to test its ability to provide the right physical insight,
we focus on the following aspects: i) the overall effect of dephasing: ii) the quantitative changes in the emission of TDI, as a function of the dephasing time $T_2$: iii) the effect of the detuning $\delta$ { and of the intensity}; iv) quantifying coherence. { Unless otherwise specified, all the SSE calculations have been performed with an intensity of 5x10$^3$ W/cm$^2$. }\\
\begin{figure}[tbp]
\includegraphics[width=0.8\textwidth]{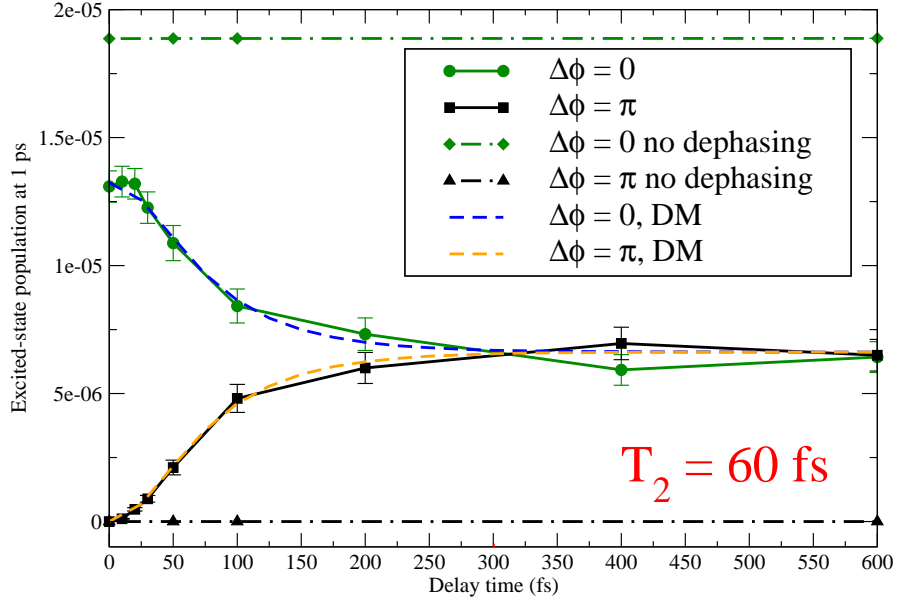}\\
\includegraphics[width=0.8\textwidth]{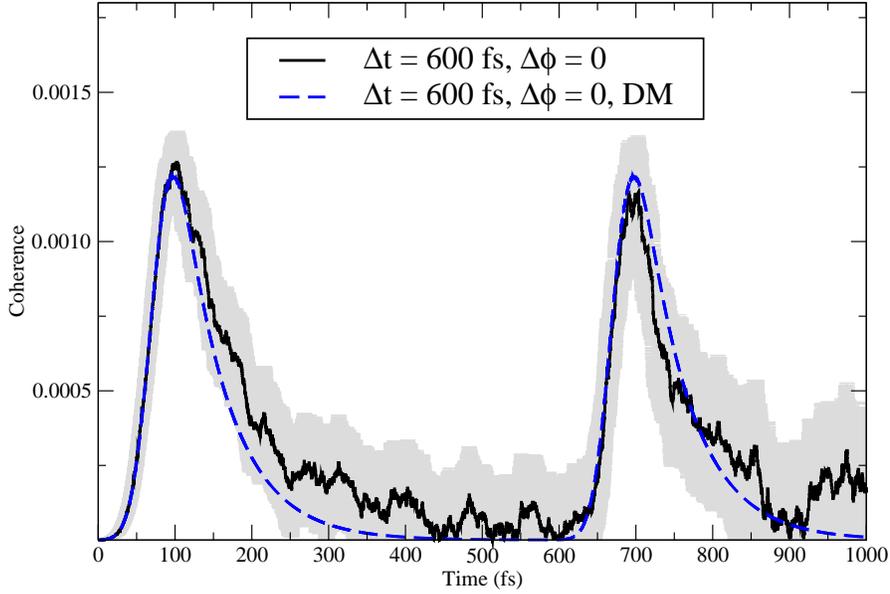}
\caption{Upper panel: excited-state population $(\hat{\rho}_S(t))_{11}$ at 1 ps as a function of the time delay $\Delta t$ (in fs) for phase shifts $\Delta \phi = 0$ and $\pi$. A dephasing time $T_2 = 60$ fs has been used. Populations obtained without dephasing are also reported for a comparison. Lower panel: Time evolution of the  coherence $(\hat{\rho}_S(t))_{10}$ for $\Delta t = 600$ fs and $\Delta \phi = 0$. The shaded region corresponds to the error bar. DM curves are the results from solving the master equation for the reduced matrix.\label{fig2} }
\end{figure}
We start by considering the qualitative effect of dephasing on the emission signal of TDI. In the upper panel of Figure \ref{fig2}, the population of the first excited state (at the end of the simulation, 1 ps) is reported as a function of the time delay $\Delta t$ (in fs) between the two pulses, for $\Delta \phi = 0 , \pi$ and $T_2 = 60$ fs. The nonradiative decay time is estimated to be around 3.5 ns, which makes the effects of relaxation negligible \cite{tdi}. { In this specific case, indeed, loss of phase memory is largely dominated by the pure dephasing time $T_2$ (much shorter than the relaxation one), even though in principle also relaxation plays a role in the decoherence process, as one can argue from the application of $\hat{\mathcal{L}}$ to the reduced density matrix.} \\
According to the analysis reported in Ref. \cite{tdi}, electronic coherence in TDI is interrogated by the second pulse, and gives rise to interference features, provided that the time delay $\Delta t$  is smaller than the dephasing time. In fact, the phase coherence is gradually erased by the interaction with the environment. As the values of $\Delta t$ become larger than $T_2$, the population of the excited state tends to become independent of the phase shift applied to the second pulse. In Figure \ref{fig2}, we report the population of the first excited state, for $\Delta \phi = 0$ and $\Delta \phi = \pi$. At zero delay, these two values of the phase shift give rise to constructive and destructive interference, respectively, resulting in larger excited state occupation for $\Delta\phi=0$. 
In Figure \ref{fig2} we have also reported results obtained by solving the master equation for the reduced density matrix (DM), using the same {\it ab initio} quantities (energies, transition dipole moments) of the SSE calculations. A quantitative agreement between SSE and DM data is found within the statistical error, thus validating our approach and its implementation. \\
Even though an inversion in the values of the populations is seen for a delay time of 400 fs, they are identical within the error bars. To clarify this aspect, we repeated the calculation by doubling the number of trajectories (from 512 to 1024): the gap between the population for the two cases ($\Delta\phi=0$ and $\Delta\phi=\pi$) converges to zero, which is also the DM finding. (see Figure 1 in the Supporting Information). 
We note that, if the molecule is considered isolated, i.e. if we switch off dephasing in the calculations, the excited-state population is independent of the time delay: its value corresponds to twice the value obtained for two excitations summed incoherently for $\Delta \phi = 0$, while excited-state population vanishes due to  destructive interference in the case $\Delta \phi = \pi$. \\   
These interference effects are related to the coherence between $\vert 0 \rangle$ and $\vert 1 \rangle$, whose time evolution is reported in the lower panel of Figure \ref{fig2}. The shaded region corresponds to the uncertainty produced by the average over a finite number of SSE quantum trajectories. The build up of the coherence is clearly seen at short times, namely for the first 100 fs, and is due to the pump pulse, generating a superposition of ground and excited states. At longer times, the dephasing determines a rather rapid suppression of the coherence, which eventually goes to zero (within the error bar). As explained in Section \ref{theory}, the decay is exponential with a rate equal to $\gamma_0 + \gamma_1 = 1/T_2 ${, since the contribution to decoherence due to the relaxation is negligible here}. \\
In order to get a more quantitative insight into the emission properties of the fluorophore, we have repeated the calculations with different values of the dephasing time. Figure \ref{fig3} collects time evolutions of the excited-state populations for $T_2 = 30$ fs (panel A), $T_2 = 60$ fs (panel B, same data of Figure \ref{fig2}) and $T_2 = 120$ fs (panel C). { Using three values of $T_2$ allowed us to test our approach more effectively in different dephasing regimes.} The value of $\Delta t$ corresponding to the merge of the curves with opposite phases increases with the dephasing time. In fact, if $\Delta t$ is much larger than $T_2$, the excitations induced by the two pulses sum up incoherently, and the final population of the excited state approximately coincides with twice that induced by each pulse. The excited state population for $\Delta t = 0$ fs and $\Delta \phi = 0$ is seen to increase with the dephasing time, approaching the value obtained in the absence of dephasing (upper panel of Figure \ref{fig2}) in the limit where $T_2$ is much larger than the duration of the laser pulse. This reflects the effect of dephasing already during the excitation of the system by means of a single laser pulse.
\\
In panel D of Figure \ref{fig3} we report the experimental emission for a single TDI molecule \cite{tdi} as a function of the time delay and for $\Delta \phi = 0$ and $\pi$, { together with SSE results obtained with a low (I=5x10$^3$ W/cm$^2$) and a high intensity (I=6.6x10$^8$ W/cm$^2$), within the range of values used in the experiments. SSE data have been scaled to match the value at 600 fs in the two cases and to superimpose the experimental profile for phase-independent values, since the proportionality factor between the (experimental) fluorescence count and the (computed) value of the excited-state population at 1 ps is unknown. Regarding the $\Delta \phi = 0$ curves, SSE points at high intensity are smaller for short delay times because of the occurrence of Rabi oscillations \cite{tdi}. For both intensities, a dephasing time $T_2$ of 60 fs has been used.} \\
Comparison with DM results for $T_2$ = 30 and 120 fs is shown in Figure S2 and S3 of the Supporting Information. Furthermore, Figures S4, S5 and S6 in the Supporting Information collect the SSE and DM time evolution of the population $(\hat{\rho}_S(t))_{11}$ for $T_2 = 30, 60$ and 120 fs, $\Delta t$ = 100 and 600 fs, and $\Delta \phi$ = 0, indicating that SSE and DM profiles coincide within the error bar. { Results in panels A, B and C of Figure \ref{fig3} have been obtained in resonance conditions.} \\
Comparison between the SSE reported reported in panel B of Figure \ref{fig3} (where we set $T_2 = 60$ fs) and the results in panel D of Figure \ref{fig3}  shows a nice qualitative agreement. However, we also note a large difference in the value of $\Delta t$ where the two curves meet: indeed, they meet at around 100-130 fs in the experiment, while at around 300 fs in the simulations. { SSE results in Figure \ref{fig3}D have been obtained with a detuning $\delta = 80$ cm$^{-1}$.}  

\begin{figure}[htbp!]
\includegraphics[width=0.43\textwidth]{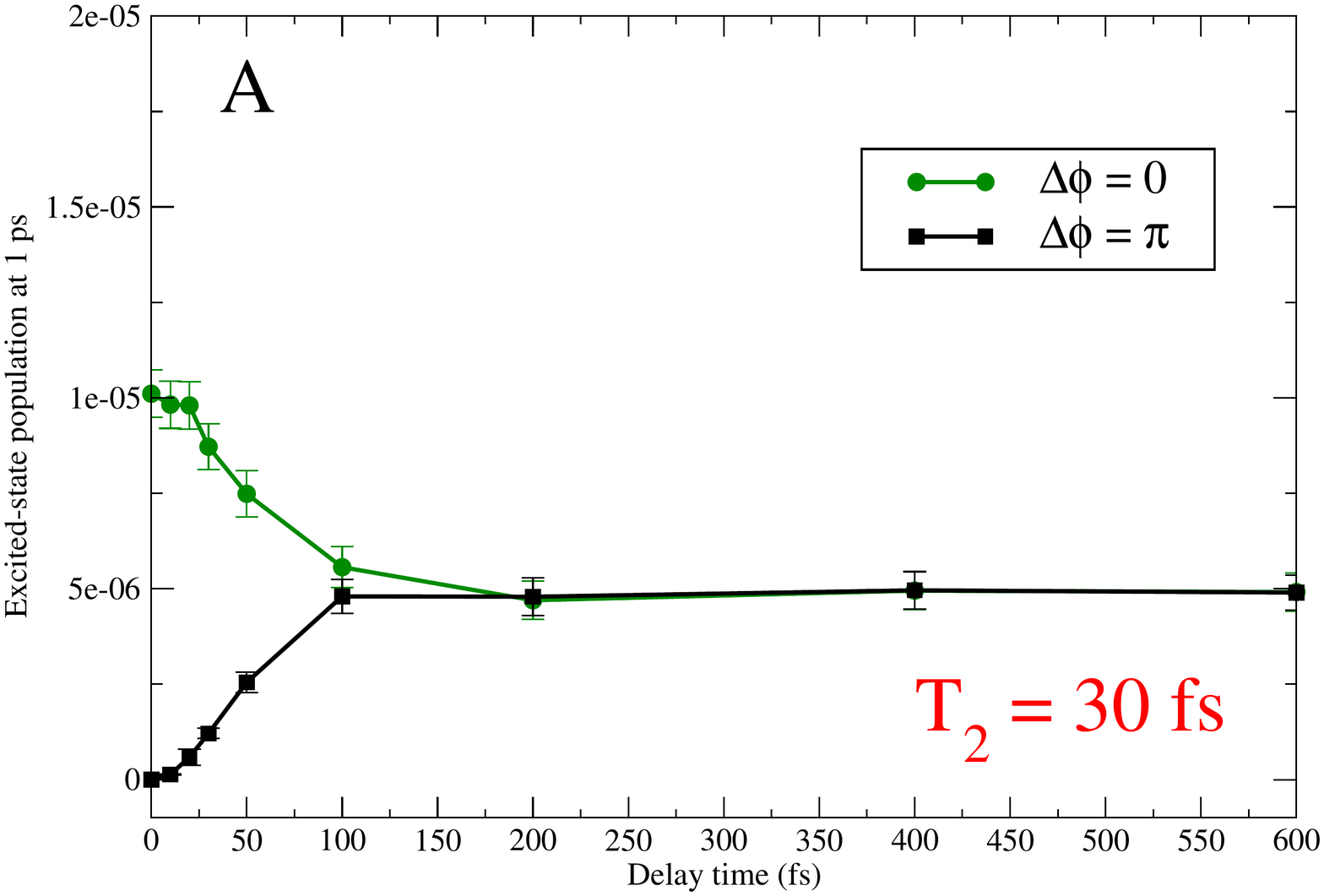}
\includegraphics[width=0.43\textwidth]{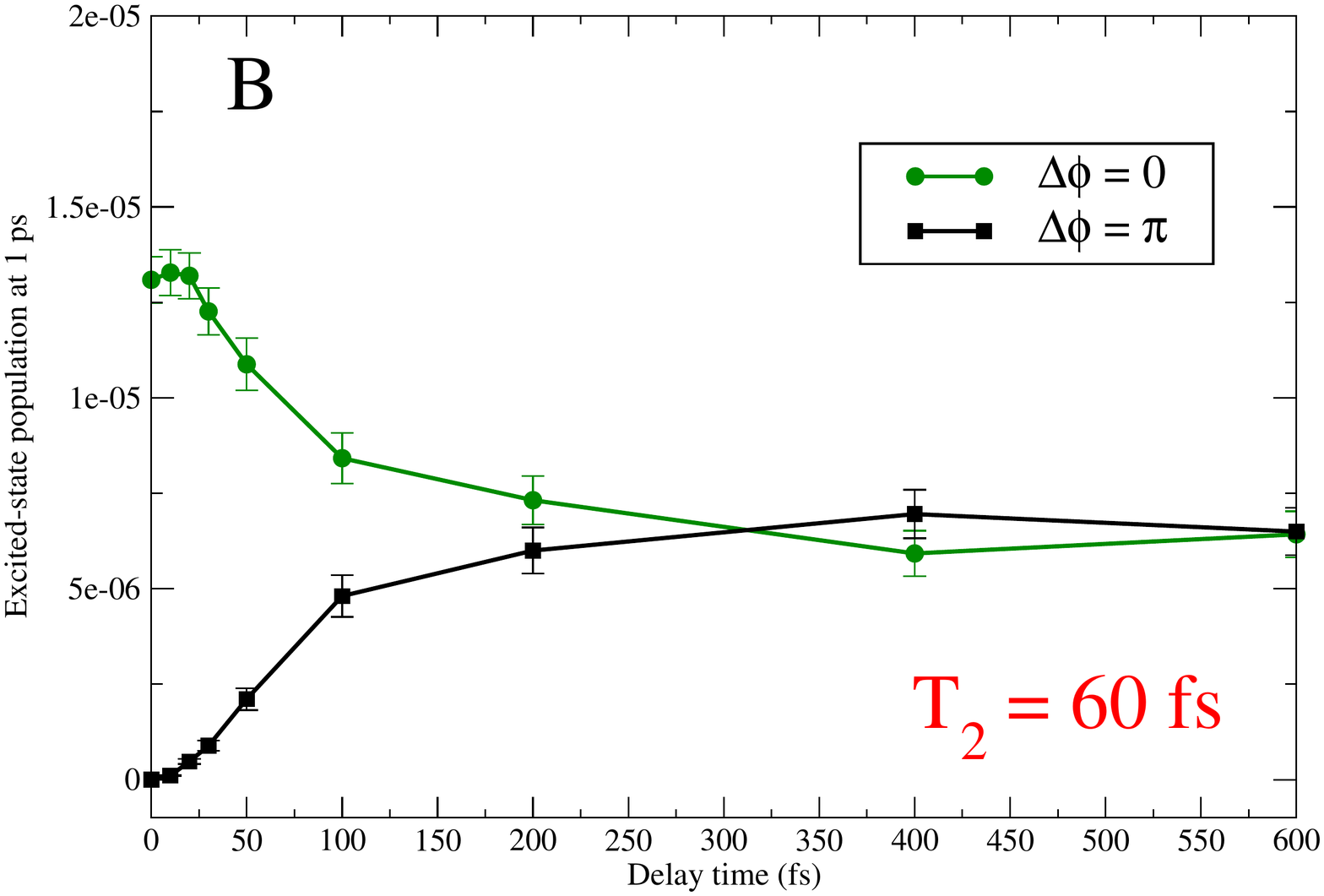} \\ 
\includegraphics[width=0.43\textwidth]{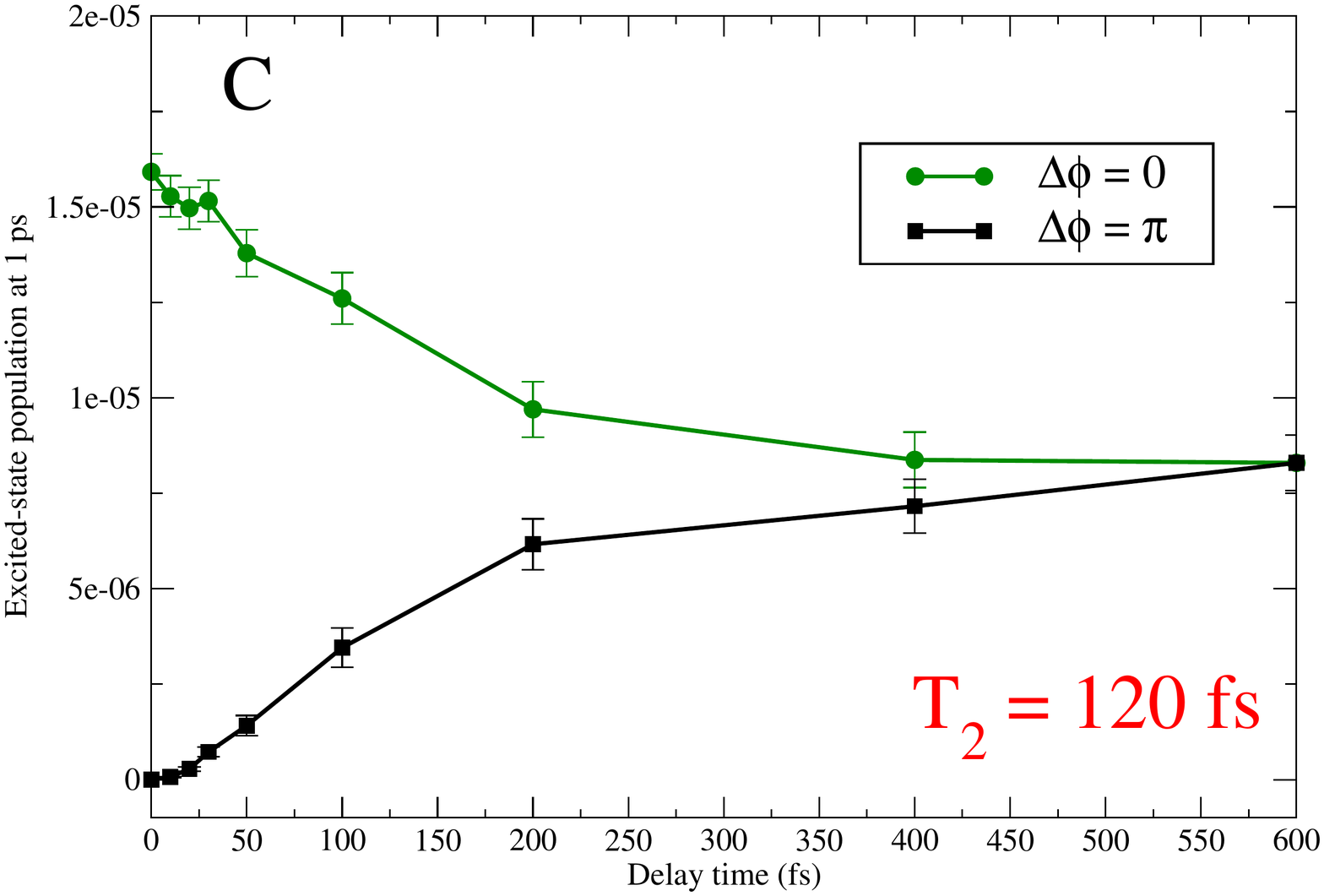}
\includegraphics[width=0.43\textwidth]{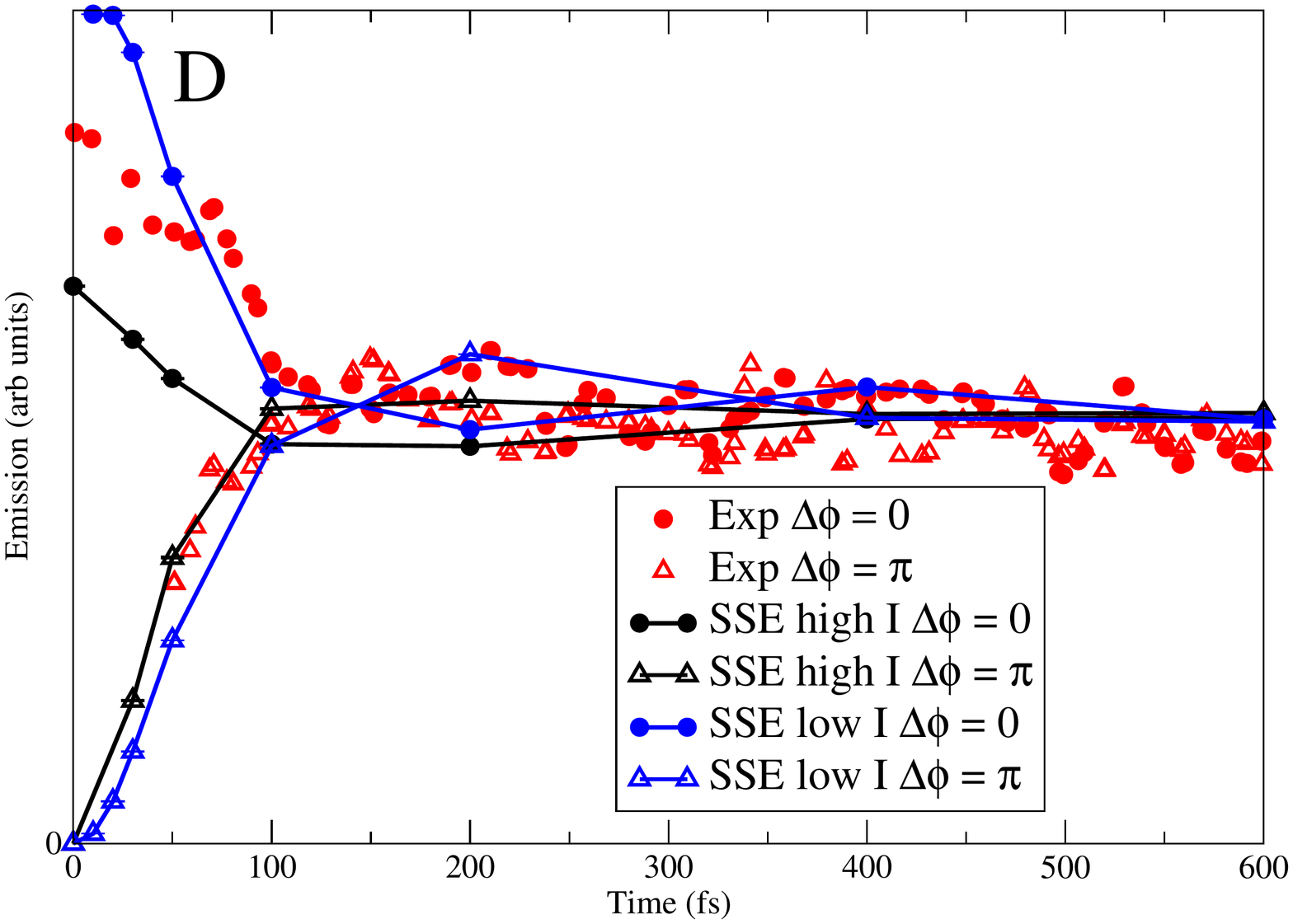} 
\caption{A: excited-state population $(\hat{\rho}_S(t))_{11}$ at 1 ps as a function of the time delay $\Delta t$ (in fs) for phase shifts $\Delta \phi = 0$ and $\pi$, and $T_2 = 30$ fs. B: same as in A but with $T_2 = 60$ fs. C: same as in A but with $T_2 = 120$ fs. D: { direct comparison between experimental data\cite{tdi} and present SSE results with I=5x10$^3$ W/cm$^2$ ("low I") and I=6.6x10$^8$ W/cm$^2$ ("high I"). Details on how panel D has been prepared are in the text.} \label{fig3} }
\end{figure}

\begin{figure}[htbp!]
\includegraphics[width=0.55\textwidth]{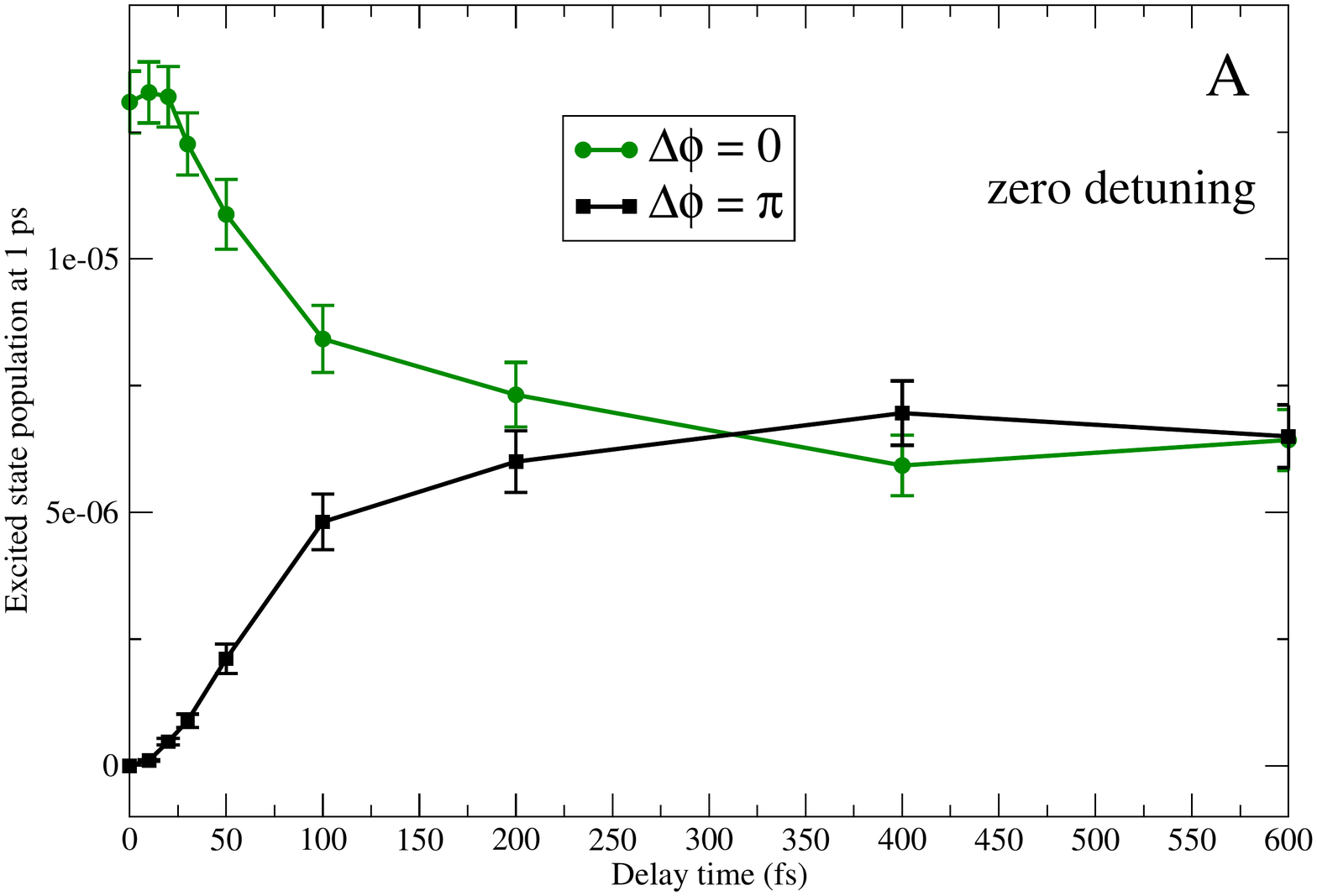} \\
\includegraphics[width=0.55\textwidth]{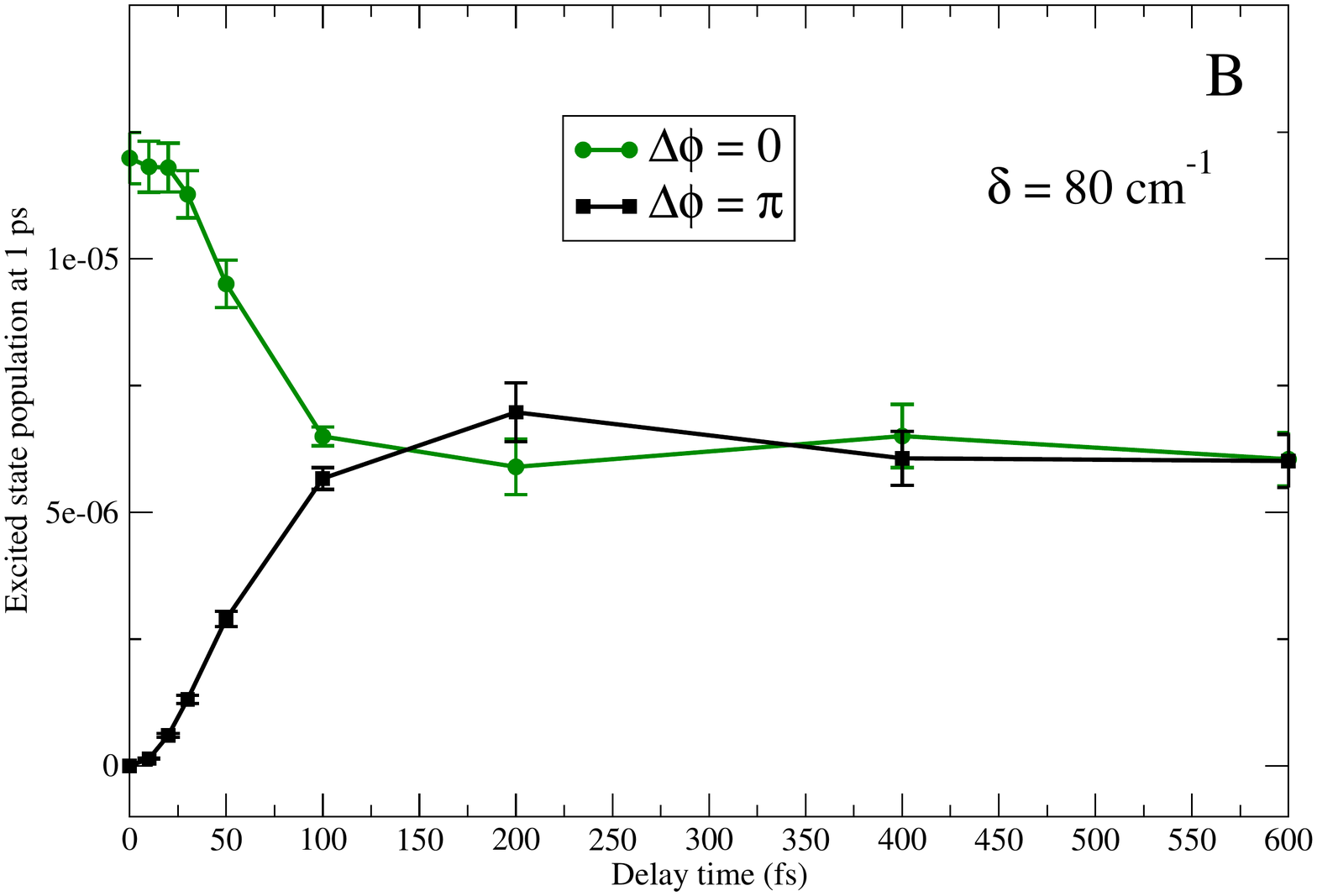} \\
\includegraphics[width=0.55\textwidth]{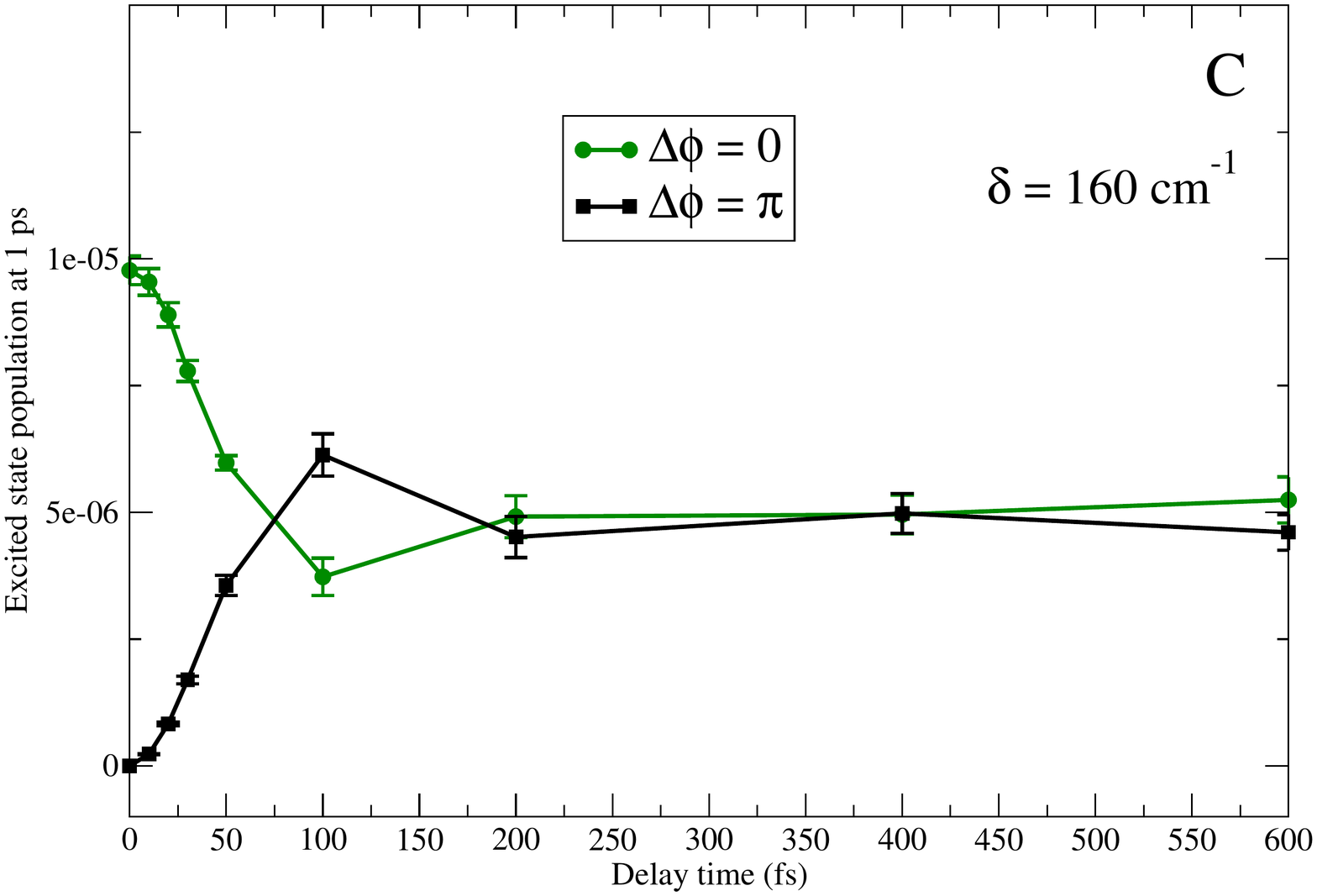}  
\caption{Excited-state population $(\hat{\rho}_S(t))_{11}$ at 1 ps as a function of the time delay $\Delta t$ (in fs) for phase shifts $\Delta \phi = 0,\pi$, $T_2 = 60$ fs, and detunings $\delta = 0$ (panel A), 80 (panel B) and 160 cm$^{-1}$ (panel C). \label{fig4} }
\end{figure}

\begin{figure}[tbp]
\includegraphics[width=0.8\textwidth]{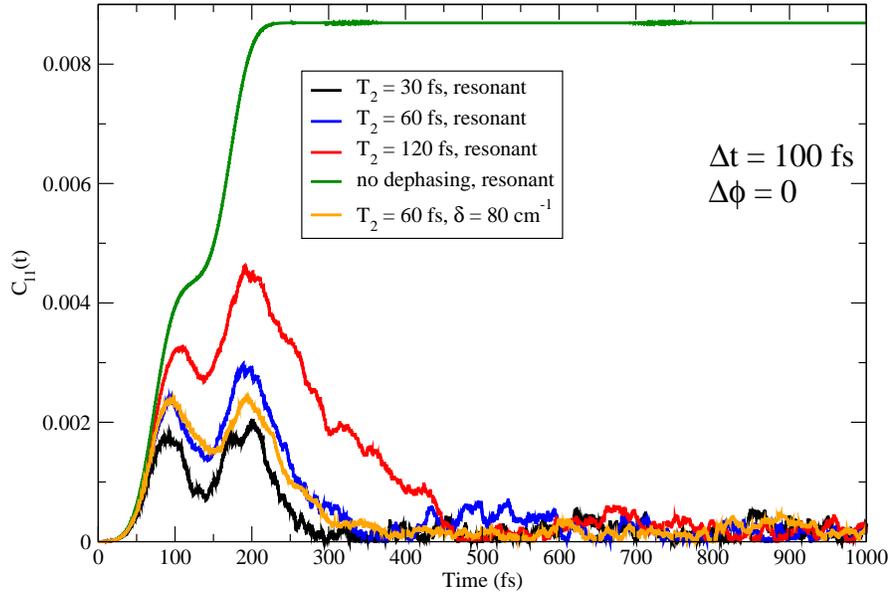}\\
\includegraphics[width=0.8\textwidth]{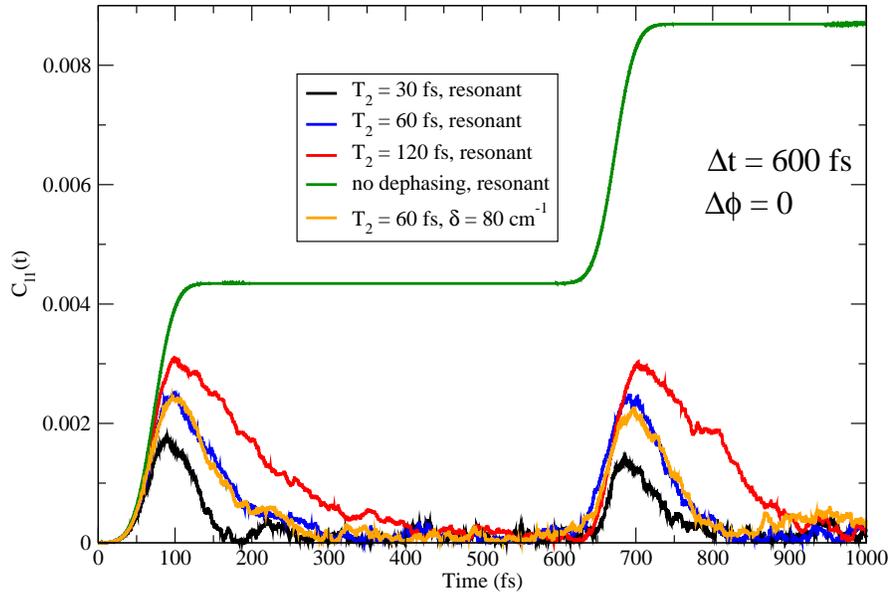}
\caption{Upper panel: $l_1$-norm coherence quantifier $C_{l1}(t)$ as a function of time, for delay time $\Delta t = 100$ fs and phase shift zero. Lower panel: $l_1$-norm coherence quantifier  $C_{l1}(t)$ as a function of time, for delay time $\Delta t = 600$ fs and phase shift zero. Error bars are not reported for sake of clarity. \label{fig5} }
\end{figure}

Indeed, in order to reproduce quantitatively the measured fluorescence, we have to take into account that the molecule in the experiment is not excited at perfect resonance, and thus we have to include a finite detuning $\delta$ in the simulations. Based on the information in Ref. \cite{tdi}, reasonable values of $\delta$ range from few cm$^{-1}$ to around 500 cm$^{-1}$. The effect of detuning is reported in Figure \ref{fig4}, with $\delta = 80$ and 160 cm$^{-1}$; populations obtained resonantly are also reported for comparison. The presence of detuning induces oscillations of the final excited state populations. Besides, the larger the detuning, the earlier a crossing between the populations corresponding to $\Delta\phi=0$ and $\Delta\phi=\pi$ is observed. 
{ Using $\delta = 80$ cm$^{-1}$}, we can reproduce the observed crossing position between the excited-state populations at around 100-150 fs { (Figure \ref{fig3}D)}. The value of 80 cm$^{-1}$ has also been extracted from the fitting procedure in Ref. \cite{tdi}. Effect of the detuning has been also investigated at DM level (Figures S7 and S8 of the Supporting Information): comparison between SSE and DM profiles shows a good agreement in reproducing oscillations and the detuning-dependent position of the first crossing between $\Delta \phi$ = 0 and $\pi$ populations. \\ 
For TDI, in the excitation regime studied in the present work, multi-state effects do not play a significant role, as verified by comparing the present results with a two-state calculation for $T_2$ = 60 fs in resonant conditions (Figure S9 { and S10} in the Supporting Information): for delay times of 100, 200 and 400 fs populations coincide within the error. \\
{ Including relaxation between excited states, i.e. going beyond the relaxation operator defined in Eq. \ref{eq:rel}, does not significantly change the emission pattern in this case, as reported in Figure S11 of the Supporting Information.} \\
The dependence of the excited-state population on the waiting time and on the phase reflects the time evolution of the coherences, especially that occurring between the two laser pulses. However, the simulation of the system dynamics allows us a more general and direct characterization of the overall quantum coherence of the system state. A quantitative analysis of such coherence is reported in Figure \ref{fig5}, where the $l_1$-norm of coherence $C_{l_1}$ (Eq. \ref{l1}) is shown as a function of time, for different values of the dephasing time $T_2$ and of the detuning $\delta$ (the time delay is $\Delta t = 100, 600$ fs, respectively in the upper and lower panels, and $\Delta\phi=0$). In the absence of dephasing (green curves) $C_{l_1}$ is increased by an equal amount by each laser pulse, and its final value is independent on the time delay $\Delta t$. Coherence decreases with decreasing dephasing times $T_2$. In particular, values of $T_2$ that are comparable to, or smaller than the pulse duration result in a reduction of the coherence generated by each pulse. Besides, if the phase memory is longer than the time delay (upper panel), the second laser pulse is followed by a maximum in $C_{l_1}$ that is higher than that produced by the first pulse. This corresponds to the occurrence of constructive interference in the system excitation, and thus in the observed fluorescence. If instead $\Delta t \gg T_2$ (lower panel), the coherence generated by the first pulse is completely suppressed by dephasing before the arrival of the second pulse. As a result, the amount of coherence generated by the two (identical) pulses coincides, up to fluctuations related to the averaging over the different trajectories, and no interference shows up in the final occupation of the excited state. 
\\
We finally note that the close correspondence that we have established between the dependences on $\Delta t$ and $T_2$ of the fluorescence on the one hand and of the state coherence on the other results from the fact that both are essentially related to the same, single excited state. 
In general, interference features in the molecule emission result specifically from the coherence between the ground state and the excited states that are involved in the radiative recombination, while $C_{l_1}$ accounts for the coherence between any two eigenstates.

\begin{figure}[tbp]
\includegraphics[width=0.8\textwidth]{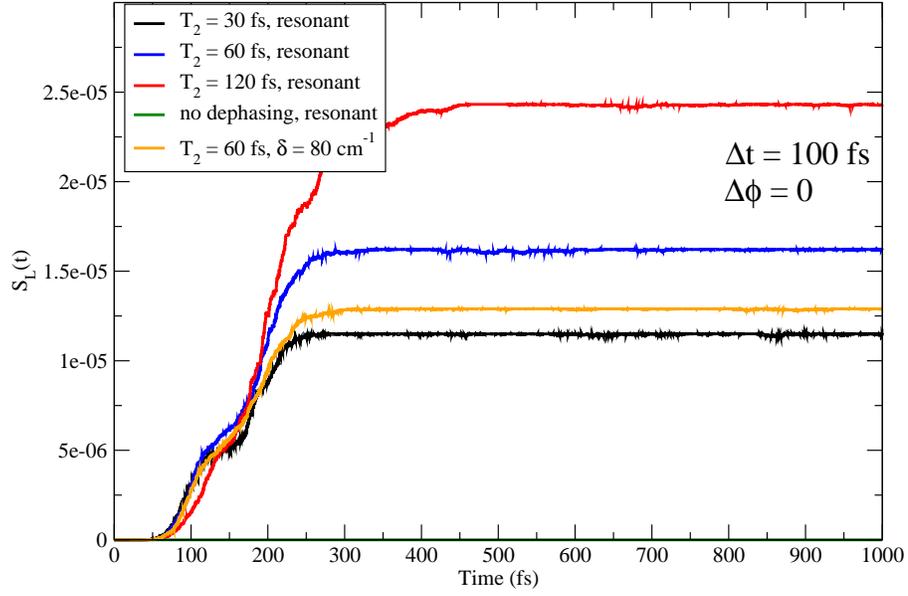}\\
\includegraphics[width=0.8\textwidth]{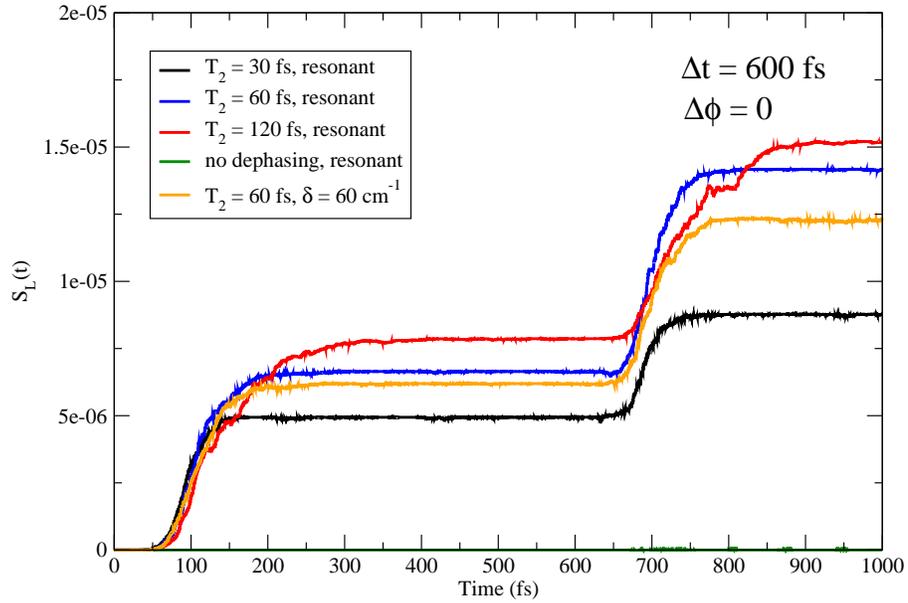}
\caption{Upper panel: linear entropy $S_L(t)$ as a function of time, for delay time $\Delta t = 100$ fs and phase shift zero. Lower panel: linear entropy $S_L(t)$ as a function of time, for delay time $\Delta t = 600$ fs and phase shift zero. Error bars are not reported for sake of clarity. \label{fig6} }
\end{figure}

 Time evolution of the linear entropy $S_L$ for the same cases as for the $l_1$-norm of coherence is reported in Figure \ref{fig6}a for a delay time of 100 fs, and in Figure \ref{fig6}b for a delay time of 400 fs. Phase shift is set to zero in both cases. \\
As mentioned above, linear entropy is a quantitative measure of the purity of the state. If the system is intialized in the ground state (or in any pure state) and in the absence of dephasing, $\rho_S$ remains a pure state throughout its time evolution, and $S_L$ is always zero.  In the presence of dephasing, the linear entropy evolves as follows. Each of the two laser pulses tends to populate the excited states and to create a linear superposition between these and the ground state. Dephasing tends to turn such a linear superposition into a statistical mixture. The asymptotic value of the linear entropy is an increasing function of the excited state population. This explains why, rather counterintuitively, the highest values of the entropy are obtained for the larger values of the dephasing time $T_2$.
In fact, with a slow dephasing (large $T_2$) populating  the excited states is more efficient: this contributes to the "disorder" of the probability distribution in the final occupations, eventually increasing $S_L$. At fixed dephasing time, $T_2 = 60$ fs, detuning makes excitation less efficient, resulting in a smaller value of the linear entropy. \\
Error bars for both $C_{l_1}$ and $S_L$ (not shown) are, along the dynamics, at least one order of magnitude smaller than the corresponding value.

\section{Conclusions}
\label{con}

We have presented a new approach, which combines SSE and quantum chemical description of a fluorophore to include dephasing and relaxation, thus enabling the study of ultrafast processess on an ab initio footing. The proposed model combines standard quantum chemistry treatment of the molecular target with the study of ultrafast processes occurring at the femtosecond scale, with the goal to study possible long-living coherence effects in time-dependent molecular properties. Relaxation and dephasing operators have been defined in the space of CIS time-independent eigenstates of the fluorophore. A key point of the proposed approach is the use of an ab initio formulation of the electronic problem related to the molecular target.\\
As a test case, we have chosen to reproduce the experimentally detected emission signal of TDI, as a function of the delay time and phase shift between two pulses \cite{tdi}.
The interplay between electronic quantum coherence and emission signal of the TDI fluorophore has been theoretically investigated by means of a real-time propagation of the stochastic Schr\"{o}dinger equation, which includes effects from a surrounding environment, such as dephasing, and relaxation. Quantitative analysis of the coherence has been also performed, using the $l_1$-norm of coherence and linear entropy.\\
We have analyzed effects of the dephasing time, the detuning and of considering more than two states. Coherence has been quantitatively investigated.
The approach was able to correctly reproduce the experimental behavior reported in Ref. \cite{tdi}. \\
Perspectives for the future work will move along three main research lines: i) the investigation of vibrational coherence by including the vibronic structure in our approach; ii) inclusion of the effects of solvent and/or nanostructures; iii) extending the method to a non-Markovian formulation of SSE within the quantum  chemistry framework. \\
i) 
Only by including vibrational structure in our model, i.e. by going beyond the pure electronic transition, we could investigate both electronic and vibrational (i.e, vibronic) coherence. For example, recently a density-matrix approach with a single vibrational mode has been applied to theoretically reproduce the vibrational modulations of emission intensity of the DN-QDI fluorophore \cite{pal17a,pal17b}.  \\
ii) The proposed model already includes the option to treat the time evolution of molecular properties embedded in a solvent and/or in presence of a nanostructure, using a polarizable continuum model \cite{pcm,tom02,car6a,cor01,and04,car06,vuk09}. 
 \\
iii)
Our aim is to extend our approach to the non-Markovian SSE \cite{gas99,bre99,sto02,rod09,rod11,rod12,sue14,bie14,rit15,veg17,bk:cohen}, in the form given in Ref. \cite{gas99}, using the polarizable continuum model, to set the proper environment response functions and fluctuations. \\
In conclusion, the presented approach represents the first promising step of a long term research line that aims at integrating all the aspects of the time evolution of molecular systems probed by ultrafast spectroscopy in a complex environment, into a {\it ab initio} based simulation framework.


\begin{acknowledgement}

Funding from the ERC under the grant ERC-CoG-2015 No. 681285 "TAME-Plasmons" is gratefully acknowledged. Computational support from HPC Lab of University of Modena and Reggio Emilia and from CINECA (Iscra C project "QNANO") is also acknowledged. Marta Rosa is gratefully acknowledged for some tests on time-propagation calculations. 

\end{acknowledgement}


\section*{Supporting Information}
{ CIS and CIS(D) excitation energies in Table S1;} comparison between results with 512 and 1024 SSE trajectories and DM data for the case reported in Figure 2 (Figure S1); comparison between SSE and DM results (Figures S2-S8); comparison between multi- and two-state calculations for $T_2$ = 60 fs (Figures S9 and S10); { comparison between SSE results using only the relaxation to the ground state (Eq. \ref{eq:rel}) or also relaxation between excited states (Figure S11). }


\bibliography{bib}

\end{document}


\begin{table}[tbp]
\tabcolsep=0.01\textwidth
\caption{Excitation energies (eV) at CIS and CIS(D) level of theory. Experimental excitation (in hexadecane) \cite{mai97} is also reported for comparison. State ordering from the CIS(D) calculation.}
\label{tab1}
\begin{tabular}{c|cc|c}
  Excited state & CIS  &  CIS(D)  & Exp      \\  \hline
1 & 2.759    &   2.474   & 1.928    \\ 
2 & 5.009    &   3.171   &     \\
3 & 5.310    &   3.352   &     \\ 
4 & 4.476    &   3.530   &     \\ 
5 & 4.222    &   3.718   &     \\ 
6 & 4.683    &   3.728   &     \\ 
7 & 4.828    &   3.871   &     \\ 
8 & 4.191    &   3.879   &     \\  
9 & 5.247    &   4.362   &     \\ 
10 & 5.334    &  4.958    &     \\  
\end{tabular}
\end{table}

\begin{figure}[tbp]
\begin{center}
\includegraphics[width=0.95\textwidth]{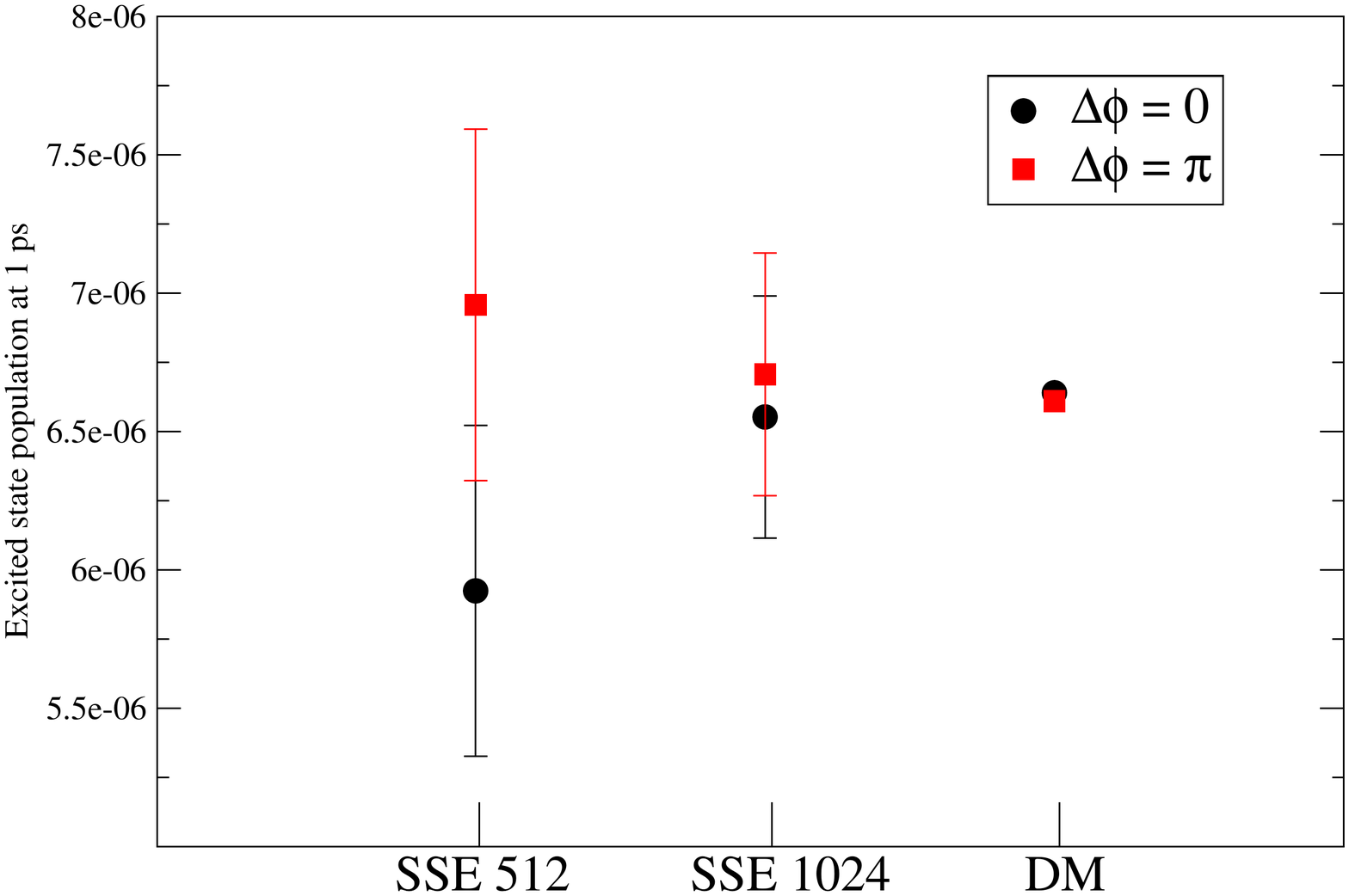}
\end{center}
\caption{Population of the first excited state of TDI for phase shift $\Delta \phi$ = 0 and $\pi$, $\Delta t$ = 400 fs and $T_2$ = 60 fs. Comparison between the results obtained with 512 and 1024 SSE trajectories, and with the density matrix (DM) approach. I = 5x10$^3$ W/cm$^2$.\label{figs1} }
\end{figure}

\begin{figure}[tbp]
\includegraphics[width=0.95\textwidth]{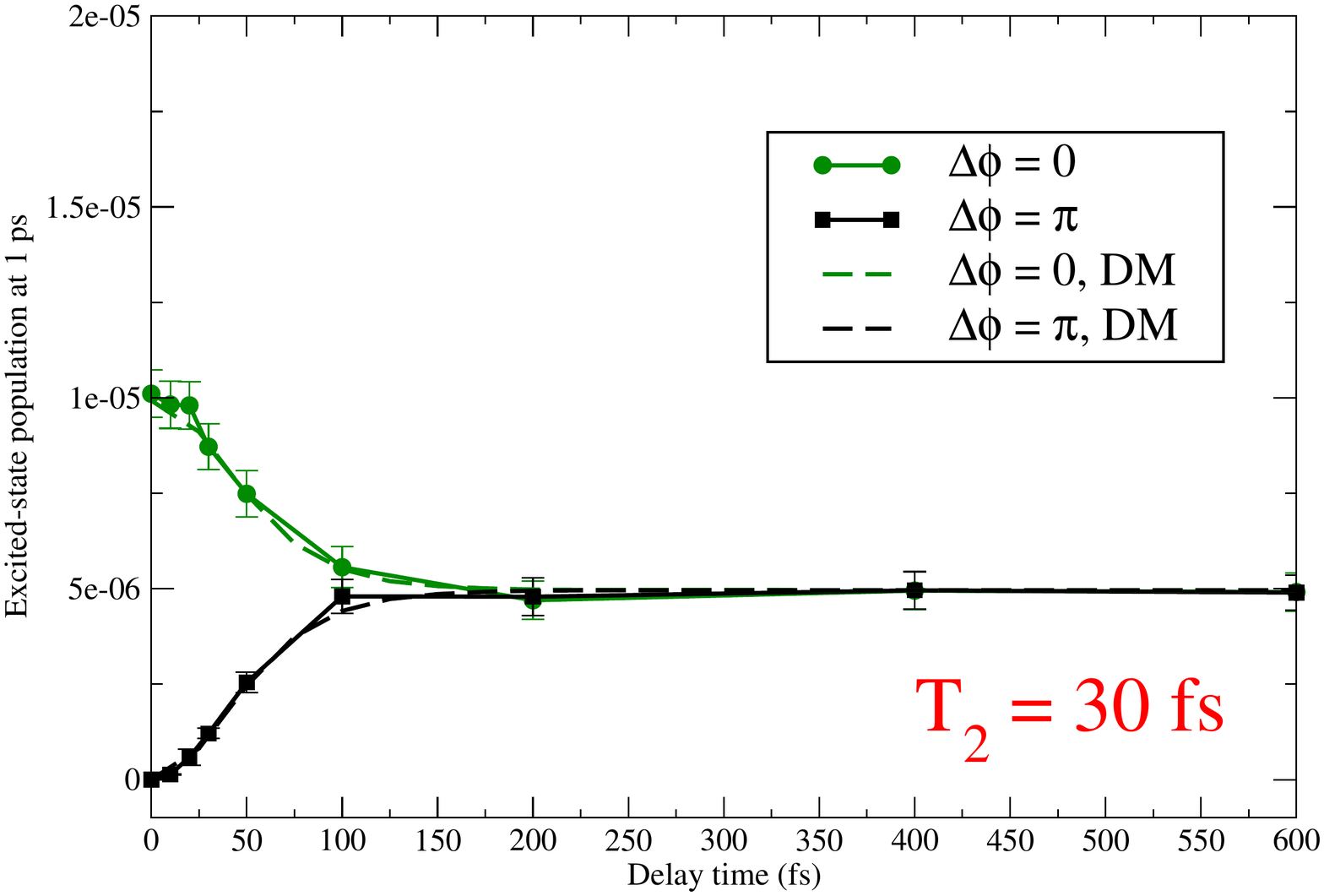}
\caption{Comparison between the population (at 1 ps) of the first excited state of TDI, between SSE (512 trajectories) and DM, as a function of the delay time (in fs). Dephasing time of $T_2$ = 30 fs and no detuning. I = 5x10$^3$ W/cm$^2$. \label{figs2} }
\end{figure}

\begin{figure}[tbp]
\includegraphics[width=0.95\textwidth]{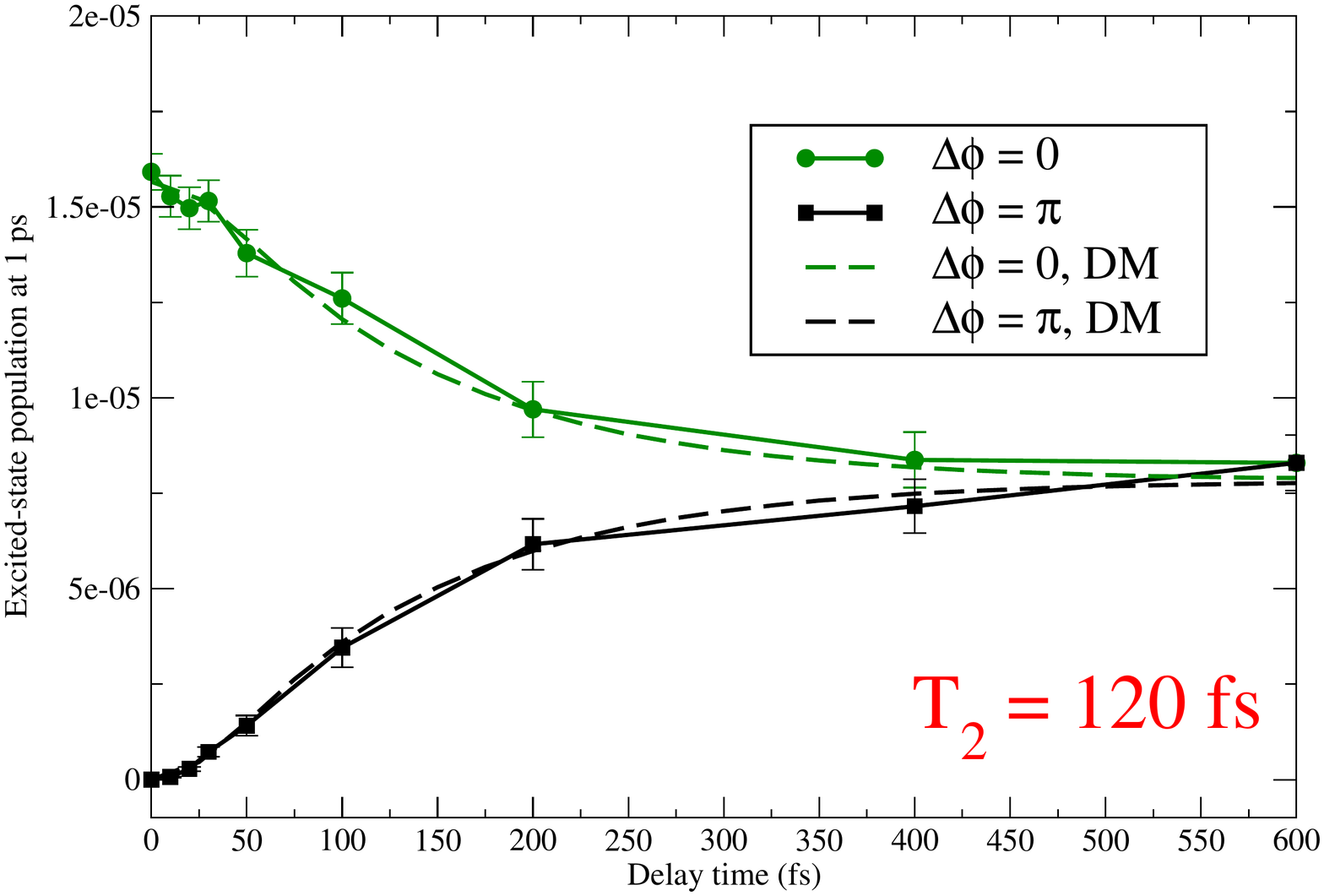}
\caption{Comparison between the population (at 1 ps) of the first excited state of TDI, between SSE (512 trajectories) and DM, as a function of the delay time (in fs). Dephasing time of $T_2$ = 120 fs and no detuning. I = 5x10$^3$ W/cm$^2$.\label{figs3} }
\end{figure}

\begin{figure}[tbp]
\includegraphics[width=0.95\textwidth]{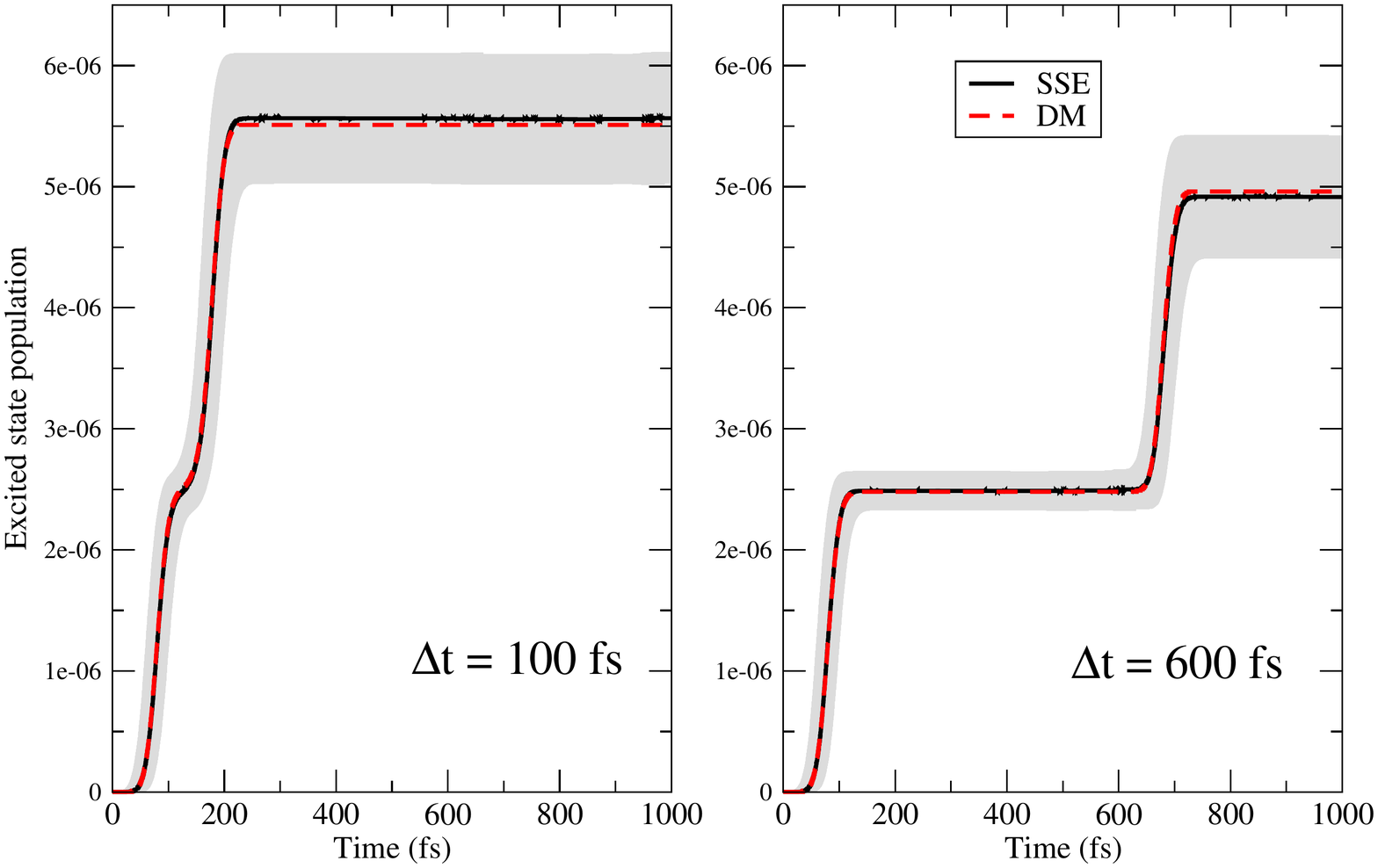}
\caption{Comparison between the time evolution of the population of the first excited state of TDI, between SSE (512 trajectories) and DM. Dephasing time of $T_2$ = 30 fs, zero phase shift and no detuning. Left panel: delay time equal to 100 fs. Right panel: delay time equal to 600 fs. I = 5x10$^3$ W/cm$^2$. \label{figs4} }
\end{figure}

\begin{figure}[tbp]
\includegraphics[width=0.95\textwidth]{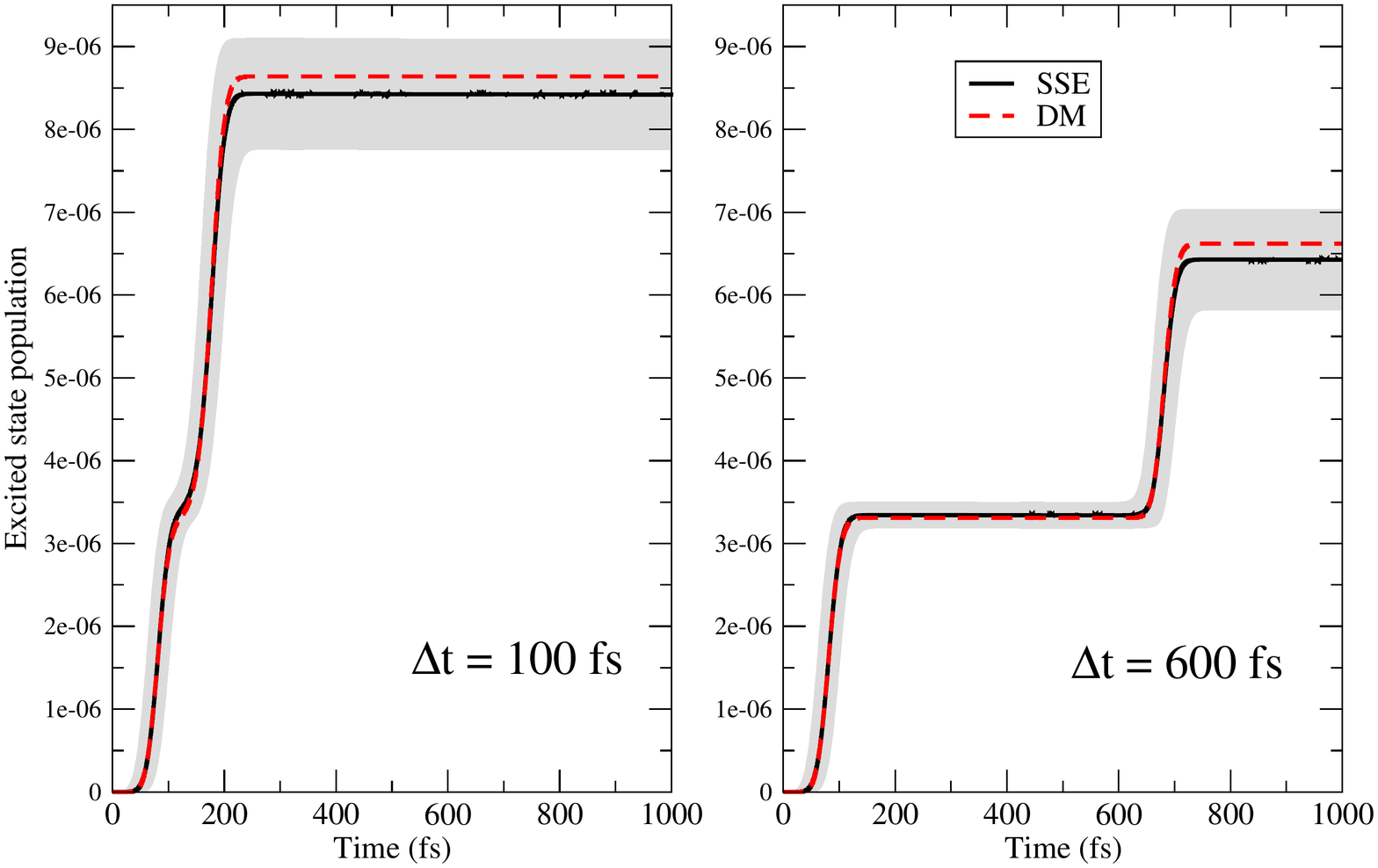}
\caption{Comparison between the time evolution of the population of the first excited state of TDI, between SSE (512 trajectories) and DM. Dephasing time of $T_2$ = 60 fs, zero phase shift and no detuning. Left panel: delay time equal to 100 fs. Right panel: delay time equal to 600 fs. I = 5x10$^3$ W/cm$^2$. \label{figs5} }
\end{figure}

\begin{figure}[tbp]
\includegraphics[width=0.95\textwidth]{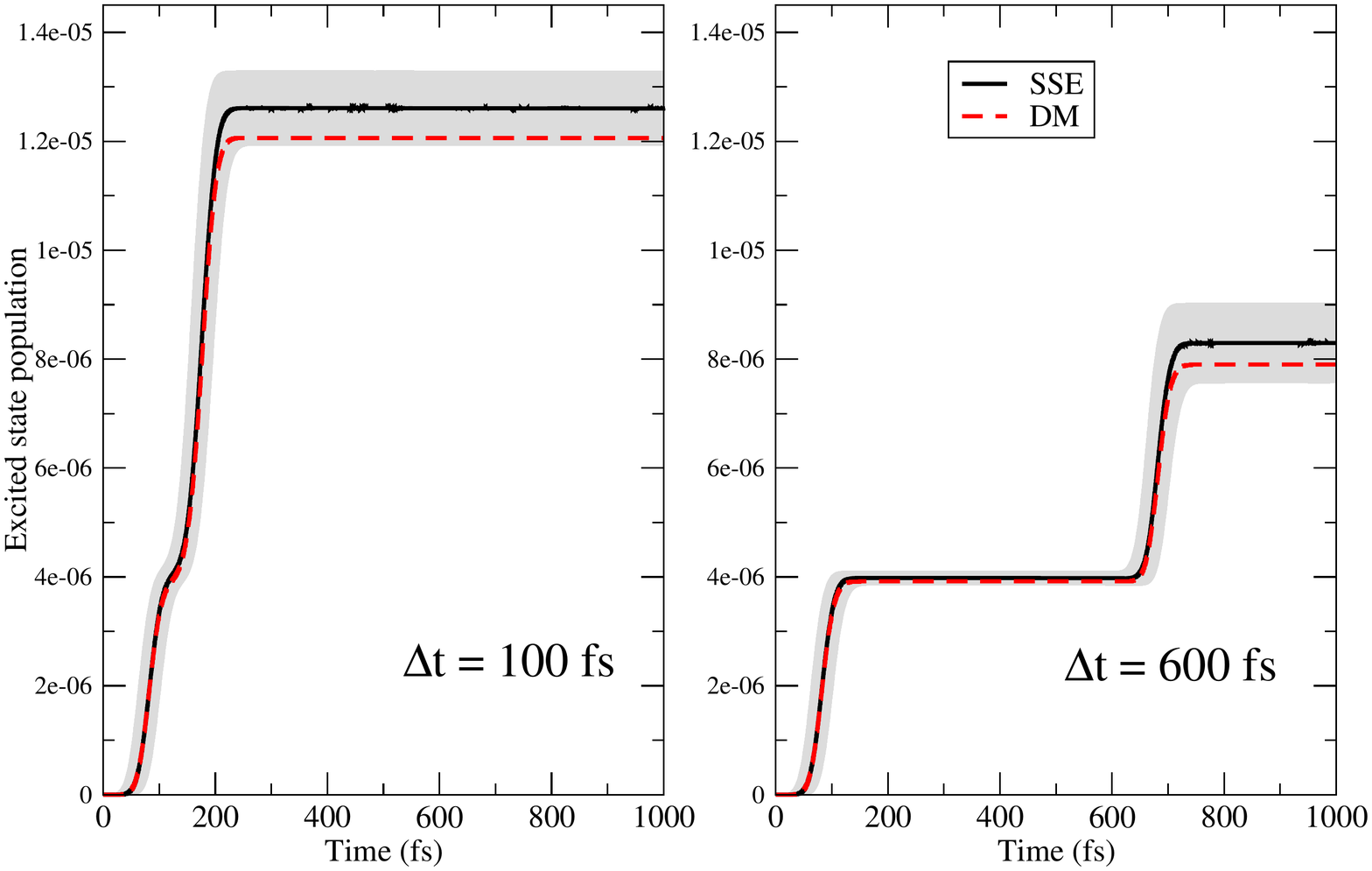}
\caption{Comparison between the time evolution of the population of the first excited state of TDI, between SSE (512 trajectories) and DM. Dephasing time of $T_2$ = 120 fs, zero phase shift and no detuning. Left panel: delay time equal to 100 fs. Right panel: delay time equal to 600 fs. I = 5x10$^3$ W/cm$^2$. \label{figs6} }
\end{figure}

\begin{figure}[tbp]
\includegraphics[width=0.95\textwidth]{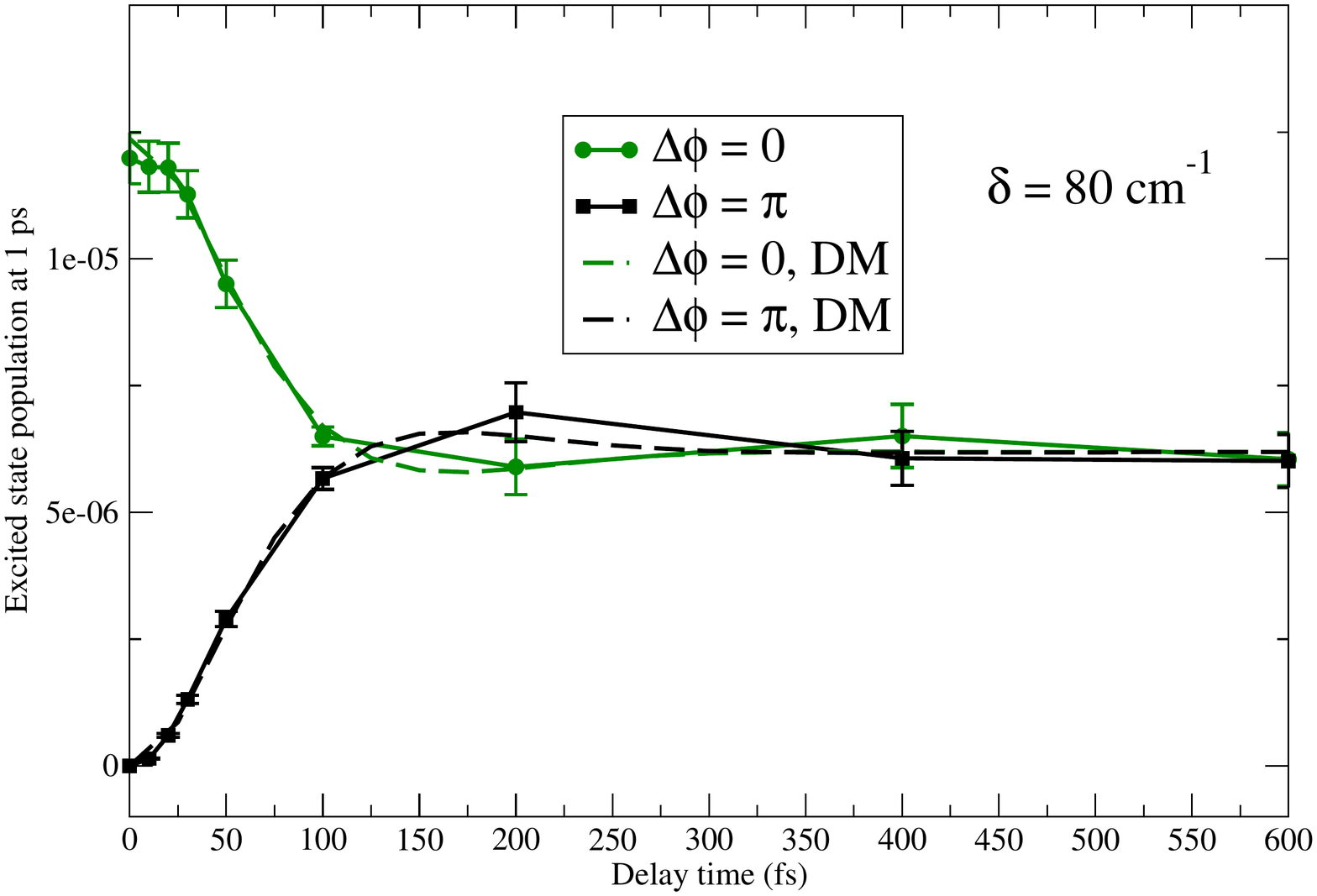}
\caption{Comparison between the population (at 1 ps) of the first excited state of TDI, between SSE (512 trajectories) and DM, as a function of the delay time (in fs). Dephasing time of $T_2$ = 60 fs and detuning $\delta$ = 80 cm$^{-1}$. I = 5x10$^3$ W/cm$^2$. \label{figs7} }
\end{figure}

\begin{figure}[tbp]
\includegraphics[width=0.95\textwidth]{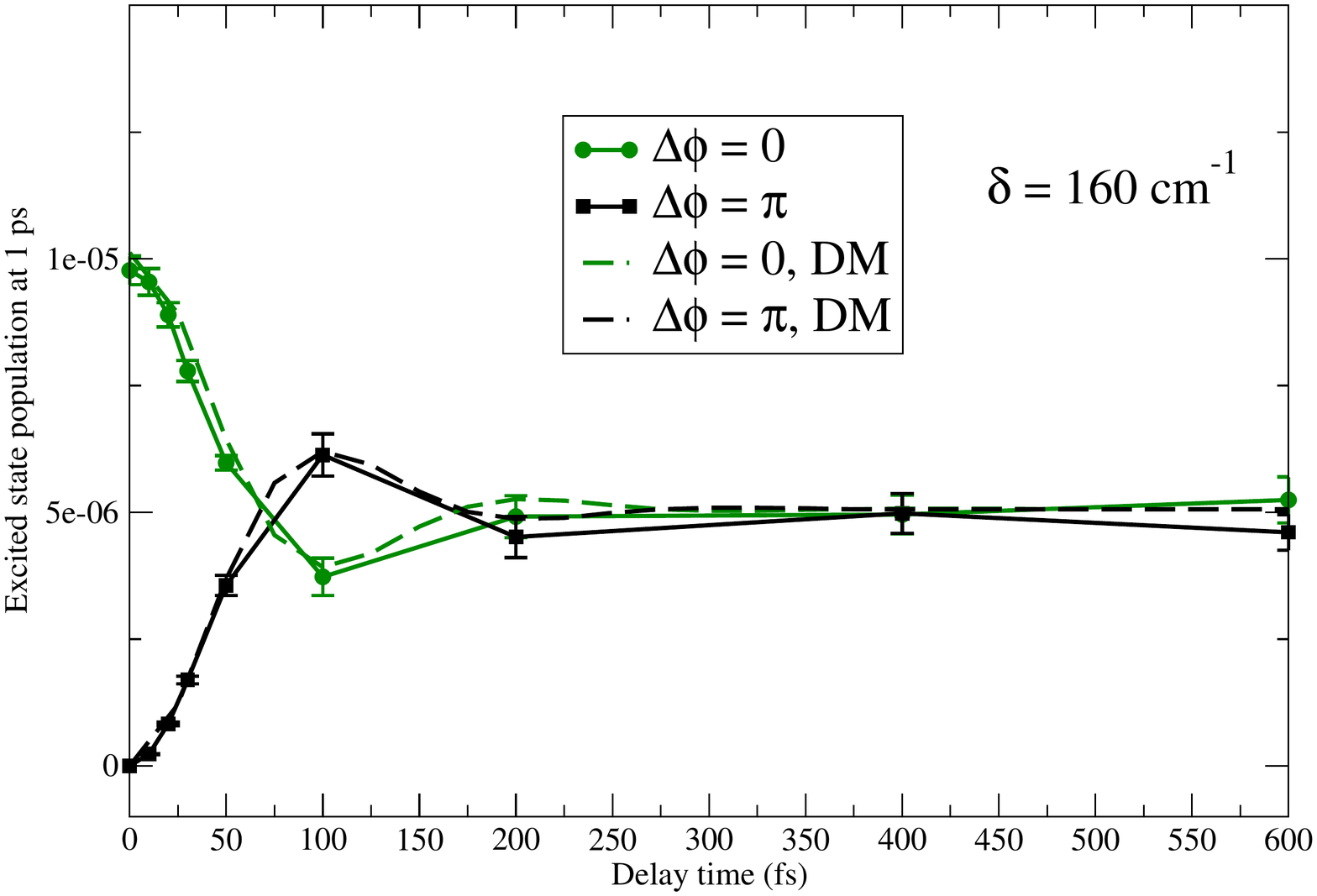}
\caption{Comparison between the population (at 1 ps) of the first excited state of TDI, between SSE (512 trajectories) and DM, as a function of the delay time (in fs). Dephasing time of $T_2$ = 60 fs and detuning $\delta$ = 160 cm$^{-1}$. I = 5x10$^3$ W/cm$^2$. \label{figs8} }
\end{figure}

\begin{figure}[tbp]
\includegraphics[width=0.95\textwidth]{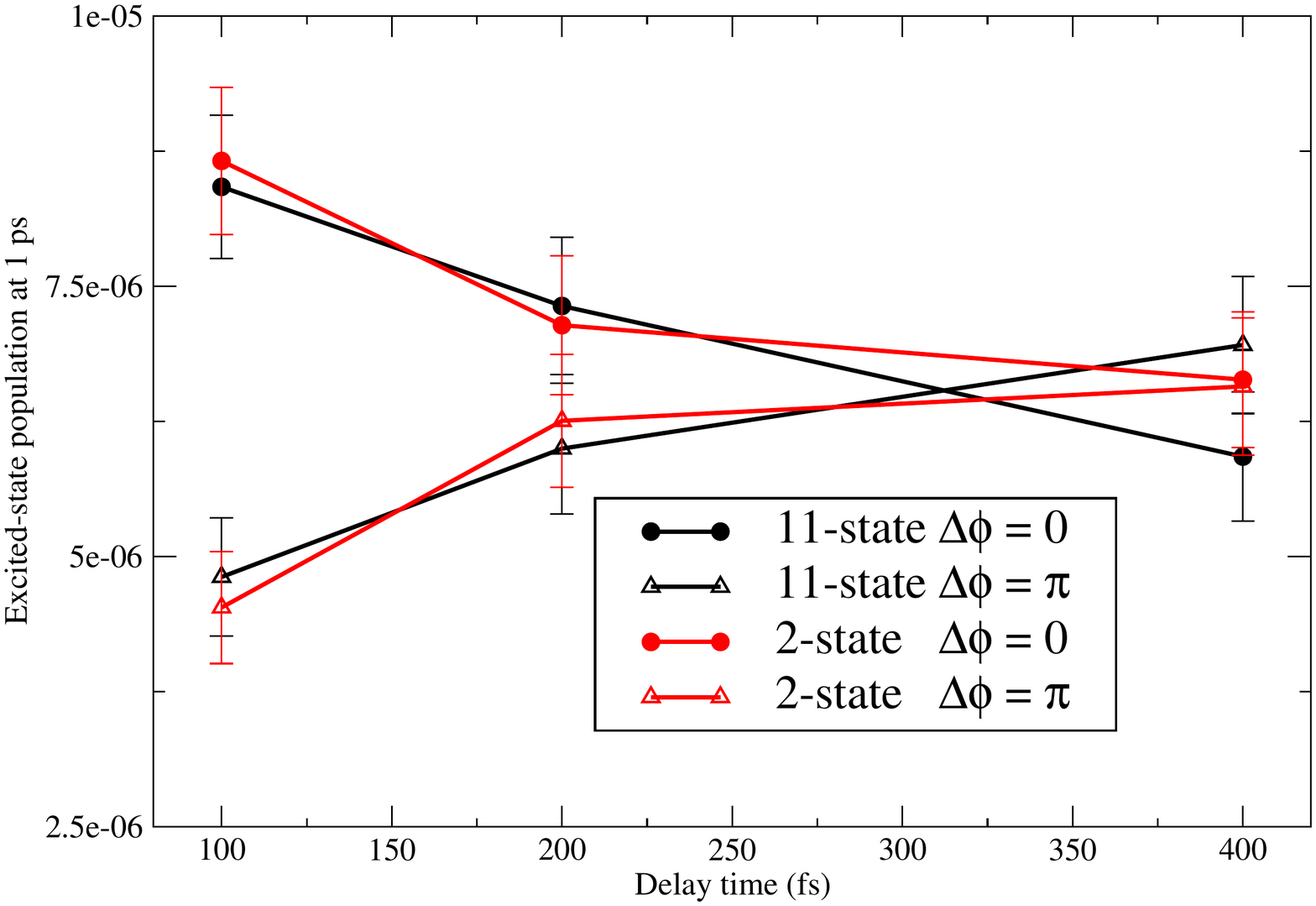}
\caption{Comparison between the population (at 1 ps) of the first excited state of TDI, from the 11-state and 2-state models, as a function of the delay time (in fs). Dephasing time of $T_2$ = 60 fs and no detuning. SSE calculations with I = 5x10$^3$ W/cm$^2$.\label{figs9} }
\end{figure}

\begin{figure}[tbp]
\includegraphics[width=0.95\textwidth]{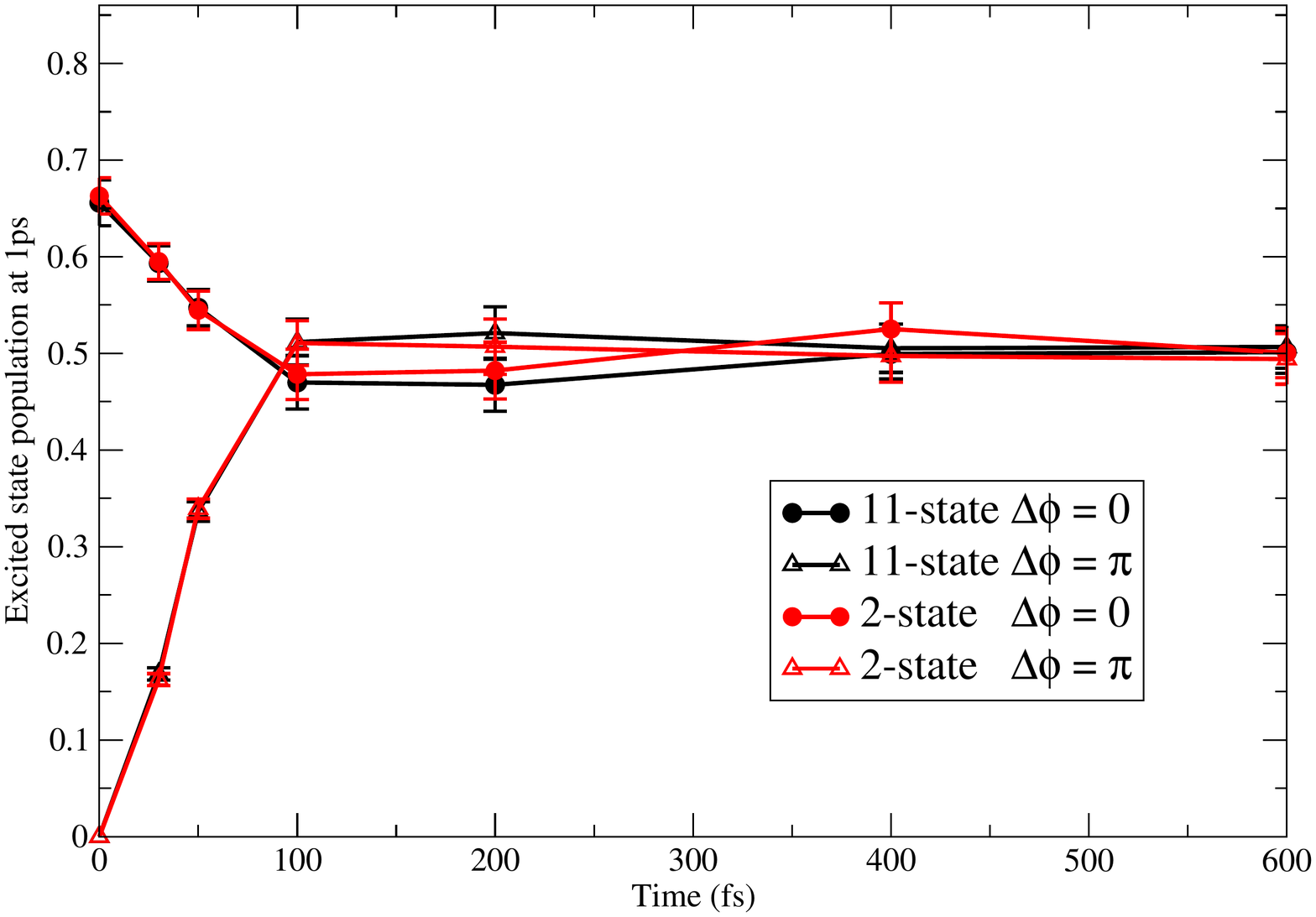}
\caption{Comparison between the population (at 1 ps) of the first excited state of TDI, from the 11-state and 2-state models, as a function of the delay time (in fs). Dephasing time of $T_2$ = 60 fs and detuning $\delta = 80$ cm$^{-1}$. SSE calculations with I = 6.6x10$^8$ W/cm$^2$.\label{figs10} }
\end{figure}

\begin{figure}[tbp]
\includegraphics[width=0.95\textwidth]{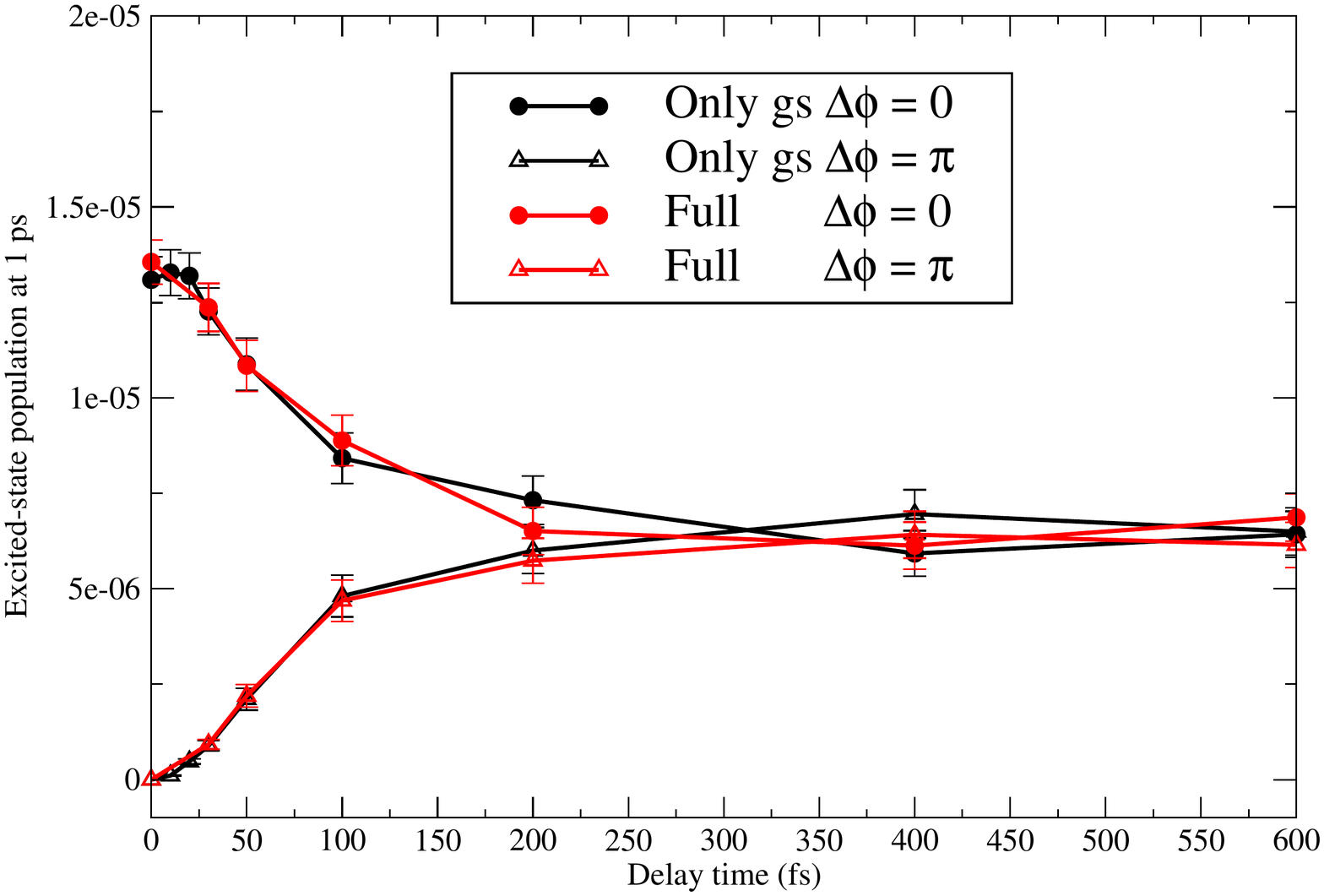}
\caption{Comparison between the population (at 1 ps) of the first excited state of TDI obtained using only relaxation to the ground state from any excited state $\vert \Phi_q \rangle$ ($\vert \Phi_q \rangle \rightarrow \vert \Phi_0 \rangle$, "Only gs") or also intermediate relaxation to the excited states ($\sum_{q>k} \vert \Phi_q \rangle \rightarrow \vert \Phi_k \rangle$, "Full"), as a function of the delay time (in fs). Dephasing time of $T_2$ = 60 fs and no detuning. SSE calculations with I = 
5x10$^3$ W/cm$^2$.\label{figs11} }
\end{figure}

\clearpage

\bibliography{bib}